\newcommand{\package}[1]{\texttt{#1}}

\documentclass[twocolumn]{aastex63}
\usepackage{amsmath}
\usepackage{amssymb}
\usepackage{euscript}
\usepackage{algorithm2e,algorithmic}
\usepackage{bbold}
\usepackage{paralist}
\usepackage{hyperref}
\usepackage{booktabs}
\usepackage{rotating}

\hypersetup{         
    unicode=true,          
    colorlinks=true,       
    linkcolor=red,          
    citecolor=blue,        
    filecolor=magenta,      
    urlcolor=red           
    }

\def\gtorder{\mathrel{\raise.3ex\hbox{$>$}\mkern-14mu
             \lower0.6ex\hbox{$\sim$}}}
\def\ltorder{\mathrel{\raise.3ex\hbox{$<$}\mkern-14mu
             \lower0.6ex\hbox{$\sim$}}}

\def\gtorder{\mathrel{\raise.3ex\hbox{$>$}\mkern-14mu
             \lower0.6ex\hbox{$\sim$}}}
\def\ltorder{\mathrel{\raise.3ex\hbox{$<$}\mkern-14mu
             \lower0.6ex\hbox{$\sim$}}}

\newcommand{\galex}{\textit{GALEX}}
\newcommand{\vph}{$v_\mathrm{ph}$}
\newcommand{\kms}{km s$^{-1}$}
\newcommand{\ergs}{erg s$^{-1}$}
\newcommand{\ek}{$E_{k}$}
\newcommand{\mej}{$M_{ej}$}
\newcommand{\msun}{$M_{\odot}$}
\newcommand{\msunyr}{$\rm M_{\odot}\ yr^{-1}$}
\newcommand{\FeII}{Fe~{\sc ii}}
\newcommand{\MgII}{Mg~{\sc ii}}
\newcommand{\Nifs}{$^{56} \rm Ni$}
\newcommand{\Cofs}{$^{56} \rm Co$}
\newcommand{\CaII}{$\rm Ca\ II$}

\newcommand{\CI}{$\rm C\ I$}

\newcommand{\ape }{SN\,2019ape}
\newcommand{\fsh }{SN\,2018fsh}
\newcommand{\uik }{SN\,2020uik}
\newcommand{\Ha}{H$\alpha$}
\newcommand{\texp}{t$_{\rm exp}$}
\newcommand{\OI}{O I}
\shortauthors{Irani et al.}

\begin{document}

\title{Less than 1\%  of Core-Collapse Supernovae in the local universe occur in elliptical galaxies}

\author[0000-0002-7996-8780]{I.~Irani}
\affiliation{Department of Particle Physics and Astrophysics,
             Weizmann Institute of Science,
             234 Herzl St, 7610001 Rehovot, Israel}
\email{idoirani@gmail.com}

\author[0000-0003-0486-6242]{S.~J.~Prentice}
\affiliation{School of Physics,
             Trinity College Dublin,
             The University of Dublin,
             Dublin 2, Ireland}
             
\author[0000-0001-6797-1889]{S.~Schulze}
\affiliation{Department of Physics, 
             The Oskar Klein Center, Stockholm University, 
             AlbaNova, SE-10691 Stockholm, Sweden}
\affiliation{Department of Particle Physics and Astrophysics,
             Weizmann Institute of Science,
             234 Herzl St, 7610001 Rehovot, Israel}

\author[0000-0002-3653-5598]{A.~Gal-Yam}
\affiliation{Department of Particle Physics and Astrophysics,
             Weizmann Institute of Science,
             234 Herzl St, 7610001 Rehovot, Israel}

\author{Jacob Teffs}
\affiliation{Astrophysics Research Institute, Liverpool John Moores University, IC2 Liverpool Science Park, 146 Brownlow Hill, Liverpool L3 5RF, UK}

\author{Paolo Mazzali}
\affiliation{Astrophysics Research Institute, Liverpool John Moores University, IC2 Liverpool Science Park, 146 Brownlow Hill, Liverpool L3 5RF, UK}
\affiliation{Max-Planck Institut fur Astrophysik, Karl-Schwarzschild-Str. 1, D-85741 Garching, Germany}

\author[0000-0003-1546-6615]{J.~Sollerman}
\affiliation{Department of Astronomy, 
             The Oskar Klein Center, Stockholm University, 
             AlbaNova, SE-10691 Stockholm, Sweden}

\author{E. ~P. ~Gonzalez}
\affiliation{Las Cumbres Observatory, 6740 Cortona Dr. Suite 102, Goleta, CA, 93117, USA}
\affiliation{Department of Physics, University of California, Santa Barbara, Santa Barbara, CA, 93106, USA}

\author[0000-0002-5748-4558]{K.~Taggart}
\affiliation{Department of Astronomy and                   Astrophysics, University of                   California, Santa Cruz, CA 95064,              USA}

\author[0000-0002-8989-0542]{Kishalay De}
\affiliation{Cahill Center for Astrophysics, California Institute of Technology, 1200 E. California Blvd. Pasadena, CA 91125, USA.}

\author[0000-0002-4223-103X]{Christoffer~Fremling}
\affiliation{Division of Physics, Mathematics and Astronomy, California Institute of Technology, Pasadena, CA 91125, USA}

\author{Daniel A. Perley}
\affiliation{Astrophysics Research Institute, Liverpool John Moores University, IC2 Liverpool Science Park, 146 Brownlow Hill, Liverpool L3 5RF, UK}

\author[0000-0002-4667-6730]{Nora L. Strotjohann}
\affiliation{Department of Particle Physics and Astrophysics,
             Weizmann Institute of Science,
             234 Herzl St, 7610001 Rehovot, Israel}

\author[0000-0002-5619-4938]{Mansi M. Kasliwal}
\affiliation{Division of Physics, Mathematics and Astronomy, California Institute of Technology, Pasadena, CA 91125, USA}

\author{A. ~Howell}
\affiliation{Las Cumbres Observatory, 6740 Cortona Dr. Suite 102, Goleta, CA, 93117, USA}
\affiliation{Department of Physics, University of California, Santa Barbara, Santa Barbara, CA, 93106, USA}

\author[0000-0002-7996-8780]{S.~Dhawan}
\affiliation{Department of Physics, 
             The Oskar Klein Center, Stockholm University, 
             AlbaNova, SE-10691 Stockholm, Sweden}
\affiliation{Kavli Institute for Cosmology, University of Cambridge, Madingley Road, Cambridge CB3 0HA, UK}

\author[0000-0003-0484-3331]{Anastasios Tzanidakis}
\affiliation{Cahill Center for Astrophysics, California Institute of Technology, 1200 E. California Blvd. Pasadena, CA 91125, USA.}

\author[0000-0002-1125-9187]{Daichi Hiramatsu}
\affiliation{Las Cumbres Observatory, 6740 Cortona Dr. Suite 102, Goleta, CA, 93117, USA}
\affiliation{Department of Physics, University of California, Santa Barbara, Santa Barbara, CA, 93106, USA}

\author[0000-0002-7252-3877]{Erik C. Kool}
\affiliation{Department of Astronomy, 
             The Oskar Klein Center, Stockholm University, 
             AlbaNova, SE-10691 Stockholm, Sweden}

\author[0000-0003-0227-3451]{J. P. Anderson}             
\affiliation{European Southern Observatory, Alonso de C\'ordova 3107, Casilla 19, Santiago, Chile}

\author[0000-0003-3939-7167]{T. E. M{\"u}ller-Bravo}
\affiliation{School of Physics and Astronomy, University of Southampton, Southampton, Hampshire, SO17 1BJ, UK}

\author[0000-0002-5884-7867]{Richard Dekany}
\affiliation{Caltech Optical Observatories, California Institute of Technology, Pasadena, CA 91125, USA}

\author[0000-0002-1650-1518]{Mariusz Gromadzki}
\affiliation{Astronomical Observatory, University of Warsaw, Al. Ujazdowskie 4, 00-478 Warszawa, Poland}

\author[0000-0003-1604-2064]{Roberta Carini}
\affiliation{INAF-Osservatorio Astronomico di Roma, Via Frascati 33, 00040 Monte Porzio Catone (RM), Italy}

\author[0000-0002-1296-6887]{L. Galbany}
\affiliation{Institute of Space Sciences (ICE, CSIC), Campus UAB, Carrer de Can Magrans, s/n, E-08193 Barcelona, Spain.}

\author{Andrew J. Drake}
\affiliation{Division of Physics, Mathematics and Astronomy, California Institute of Technology, Pasadena, CA 91125, USA}

\author[0000-0003-0035-6659]{Jamison Burke}
\affiliation{Las Cumbres Observatory, 6740 Cortona Dr. Suite 102, Goleta, CA, 93117, USA}
\affiliation{Department of Physics, University of California, Santa Barbara, Santa Barbara, CA, 93106, USA}
\author[0000-0002-7472-1279]{Craig Pellegrino}
\affiliation{Las Cumbres Observatory, 6740 Cortona Dr. Suite 102, Goleta, CA, 93117, USA}
\affiliation{Department of Physics, University of California, Santa Barbara, Santa Barbara, CA, 93106, USA}
\author{Massimo Della Valle}
\affiliation{INAF-Capodimonte Astronomical Observatory, Salita Moiariello 16, 80131 Naples, Italy}
\affiliation{INFN Naples, Naples 80126, Italy}
\affiliation{ICRANet, Piazza della Repubblica 10, I-65122 Pescara, Italy}

\author[0000-0002-7226-0659]{Michael S. Medford}
\affiliation{Department of Astronomy, University of California, Berkeley, Berkeley, CA 94720}
\affiliation{Lawrence Berkeley National Laboratory, 1 Cyclotron Rd., Berkeley, CA 94720}

\author[0000-0001-7648-4142]{Ben Rusholme}
\affiliation{IPAC, California Institute of Technology, 1200 E. California Blvd, Pasadena, CA 91125, USA}

\author[0000-0002-1229-2499]{D. R. Young}
\affiliation{Astrophysics Research Centre, School of Mathematics and Physics, Queen's University Belfast, Belfast BT7 1NN, UK}

\author[0000-0003-2375-2064]{Claudia P. Guti\'errez}
\affiliation{Finnish Centre for Astronomy with ESO (FINCA), FI-20014 University of Turku, Finland}
\affiliation{Tuorla Observatory, Department of Physics and Astronomy, FI-20014 University of Turku, Finland}

\author{Cosimo Inserra}
\affiliation{School of Physics \& Astronomy, Cardiff University, Queens Buildings, The Parade, Cardiff, CF24 3AA, UK}
\author{Rafia Omer}
\affiliation{School of Physics and Astronomy,University of Minnesota, 116 Church St SE, Minneapolis MN 55455}

\author[0000-0003-4401-0430]{David L. Shupe}
\affiliation{IPAC, California Institute of Technology, 1200 E. California
             Blvd, Pasadena, CA 91125, USA}
\author{T.-W. Chen}
\affiliation{Department of Astronomy, 
             The Oskar Klein Center, Stockholm University, 
             AlbaNova, SE-10691 Stockholm, Sweden}
\author[0000-0002-1486-3582]{Kyung Min Shin}
\affiliation{California Institute of Technology, Pasadena, CA 91125, USA}

\author[0000-0002-0301-8017]{Ofer Yaron}
\affiliation{Department of Particle Physics and Astrophysics,
             Weizmann Institute of Science,
             234 Herzl St, 7610001 Rehovot, Israel}
\author{Curtis McCully}
\affiliation{Las Cumbres Observatory, 6740 Cortona Dr. Suite 102, Goleta, CA, 93117, USA}  

\author{Matt Nicholl}
\affiliation{Birmingham Institute for Gravitational Wave Astronomy and School of Physics and Astronomy, University of Birmingham, Birmingham B15 2TT, UK}

\author{Reed Riddle}
\affiliation{Caltech Optical Observatories, California Institute of Technology, Pasadena, CA 91125, USA}
%
%

%
%
%
\begin{abstract}
We present observations of three Core-collapse supernovae (CCSNe) in elliptical hosts, detected by the Zwicky Transient Facility Bright Transient Survey (BTS). \ape\ is a SN Ic that exploded in the main body of a typical elliptical galaxy. Its properties are consistent with an explosion of a regular SN Ic progenitor. A secondary \textit{g}-band light curve peak could indicate interaction of the ejecta with circumstellar material (CSM). An \Ha-emitting source at the explosion site suggests a residual local star formation origin. \fsh\ and \uik\ are SNe II which exploded in the outskirts of elliptical galaxies. \uik\ shows typical spectra for SNe II, while \fsh\ shows a boxy nebular \Ha\ profile, a signature of CSM interaction. We combine these 3 SNe with 7 events from the literature and analyze their hosts as a sample. We present multi-wavelength photometry of the hosts, and compare this to archival photometry of all BTS hosts. Using the spectroscopically complete BTS we conclude that $0.3\%^{+0.3}_{-0.1}$ of all CCSNe occur in elliptical galaxies. We derive star-formation rates and stellar masses for the host-galaxies and compare them to the properties of other SN hosts. We show that CCSNe in ellipticals have larger physical separations from their hosts compared to SNe Ia in elliptical galaxies, and discuss implications for star-forming activity in elliptical galaxies.\\
\end{abstract}

\section{Introduction}
\label{sec:introduction}
Core-collapse supernovae (CCSNe) are widely considered to be the terminal explosion of massive ($>8\ M_{\odot}$) stars. Except for SNe Ia and Ca-Rich SNe Ib, which are likely thermonuclear explosions of white dwarf stars, all other major SN types are currently thought to have a massive star origin (\citealt{galyam2017} and references therein). Specifically, all hydrogen-rich SNe (SNe II), hydrogen-poor and silicon-poor SNe (SNe Ib/c and SNe Ibn), and superluminous SNe (SLSN) are thought to have a massive-star origin. The progenitors of most SNe II are thought to be red supergiants (RSG), as confirmed by direct progenitor detections in deep pre-explosion images (for reviews see \citealt{smartt2015a}, \citealt{vandyk2017} and references therein). The progenitors of SNe Ic have not yet been solidly detected (\citealt{Eldridge2013, smartt2015a}, but c.f. \citealt{vandyk2018}). They are thought to be either massive single Wolf-Rayet (WR) stars \citep[e.g.,][]{taddia2019}, or massive stars whose hydrogen-rich envelope has been stripped in a binary interaction \citep{Eldridge2013}. As expected, SNe Ic have been found exclusively in star-forming environments. Similarly, other CCSN types are rarely found outside of such environments (e.g. \citealt{Hakobyan2012}). \cite{suh2011} investigated previous claims of CCSNe in early-type (i.e., elliptical and lenticular) host-galaxies, but found that these were either the result of an erroneous SN classification, or that the host-galaxies had a systematically bluer ultraviolet (UV) - optical color than the early-type hosts of SNe Ia ($NUV - \textit{r}_{\rm PS1} \sim 3\ \rm mag$ compared to $NUV - \textit{r}_{\rm PS1} \sim 5.4\ \rm mag$, respectively). Recently, \cite{Sedgwick2021} reported 36 CCSNe occurring in elliptical galaxies from a sample of 421 photometrically-classified CCSNe from the SDSS-II Supernova Survey \citep{Sako2018}, and argue that elliptical galaxies account for $\sim$11\% of the cosmic star formation budget. \cite{Kaviraj2014} estimated based on the Sloan Digital Sky Survey (SDSS; \citealt{york2000}) stripe 82 data that the contribution of early-type galaxies to the cosmic star formation budget is $14\%$.

\begin{figure*}
\centering
\includegraphics[width=2\columnwidth]{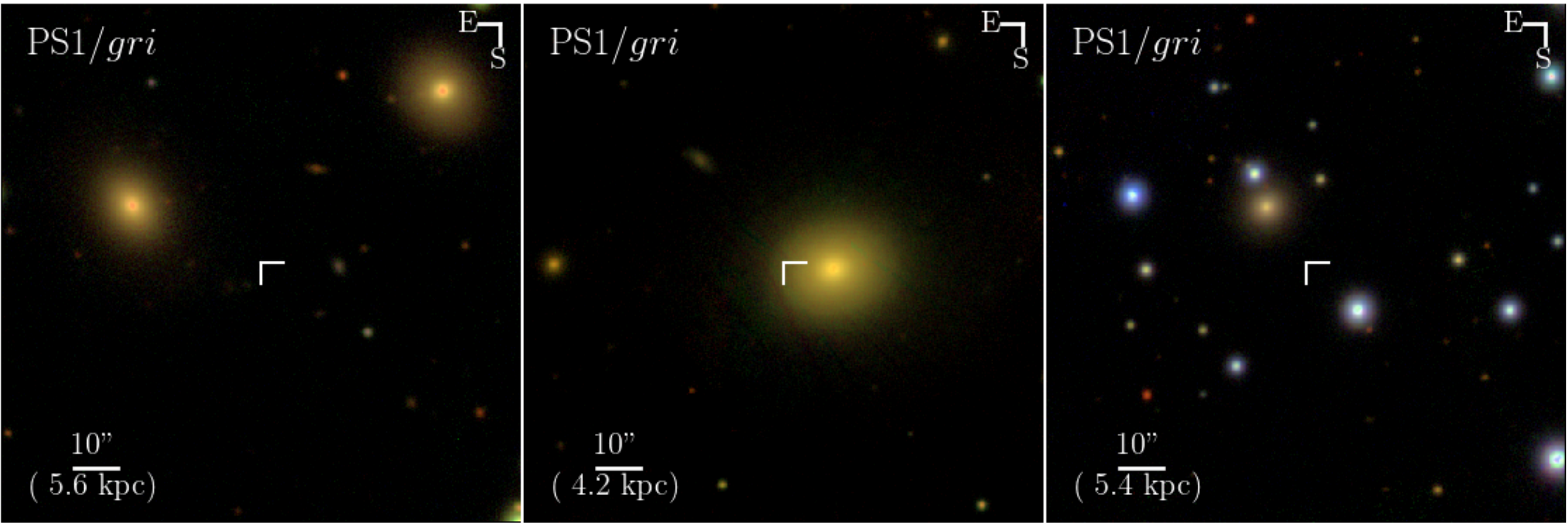}

\caption{The host galaxies of (from left to right) SN\,2018fsh, SN\,2019ape, and SN\,2020uik, constructed from PS1 stacks in the \textit{gri} bands. The location of the SN is marked with white crosshairs. The angular scale is provided in the lower left corner of each panel.}  
\label{fig:all_hosts}

\end{figure*}

\begin{figure*}
\centering
\includegraphics[width=2\columnwidth]{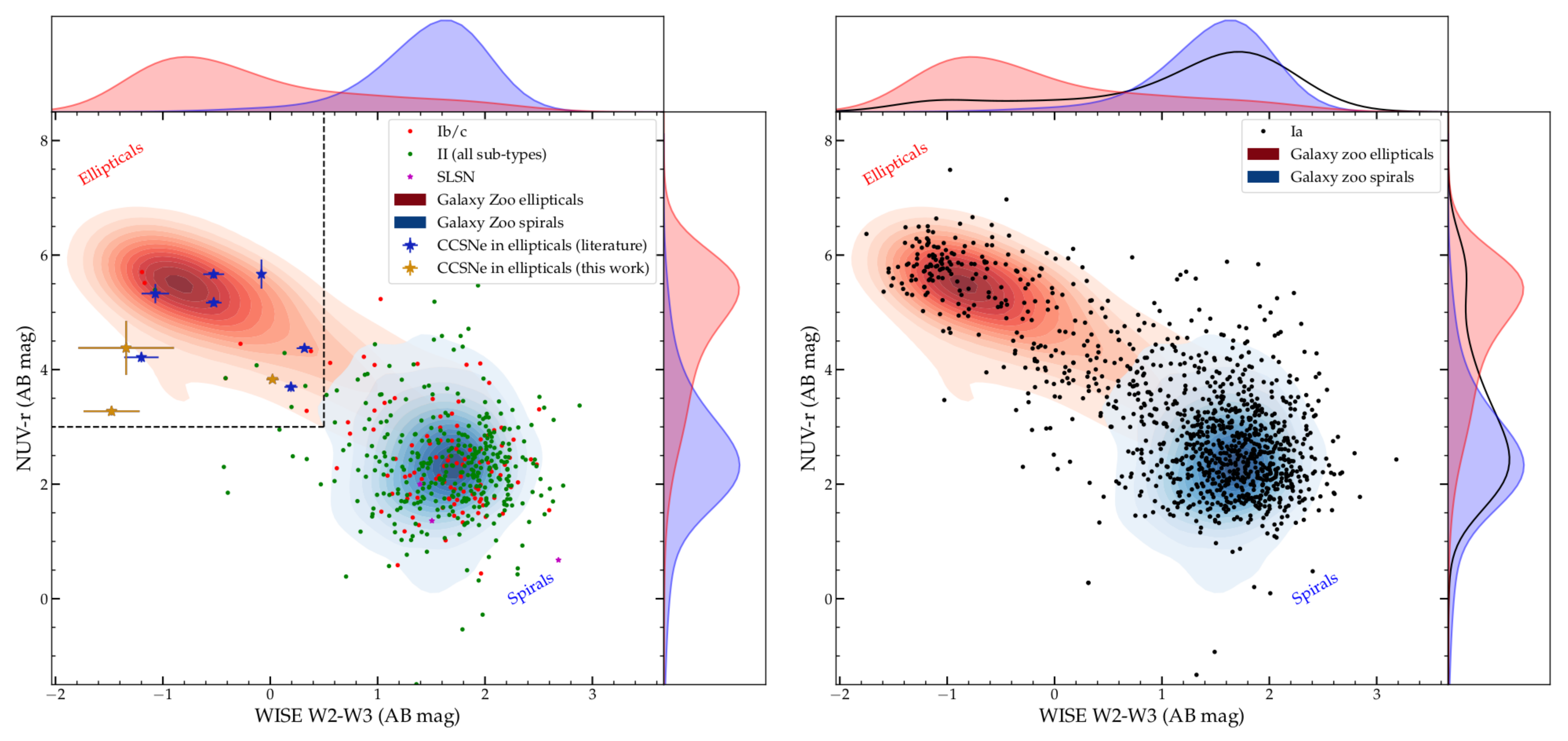}
\caption{Left: Color-color diagram for all CCSNe hosts with full color information. Each data point represents the colors of the host-galaxy of a SN. The black dashed lines indicate the region defining our color-criterion for elliptical galaxies. SNe whose hosts passed our criteria as elliptical galaxies are marked with a yellow star (this work) or a blue star (literature sample), using the host photometry reported in this work.  All BTS CCSN hosts are marked with dots color-coded based on the SN type as indicated in the legend. The red and blue shaded regions are the 2d smoothed distributions of Galaxy Zoo ellipticals and spirals. These also appear as 1d color kernel density estimate on the top and right panels. Right: Similar to the previous plot, but for the hosts of SNe Ia.}
\label{fig:color_color}
\end{figure*}    
\begin{figure*}
\centering
\includegraphics[width=1.4\columnwidth]{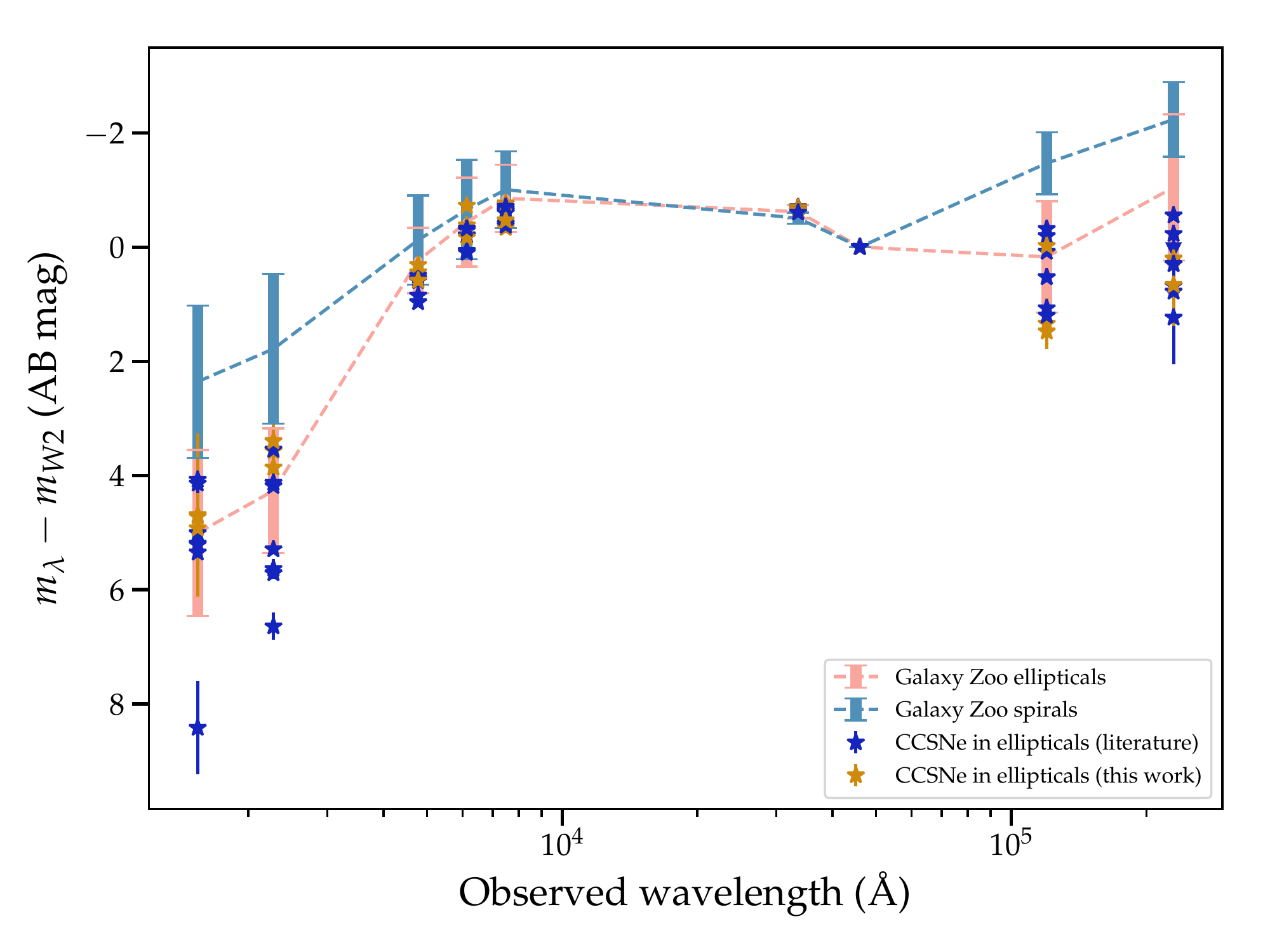}
\caption{The observed host colors relative to the W2 band for Galaxy Zoo spirals (cyan) and  Galaxy Zoo ellipticals (pink). CCSN host galaxies that passed our criteria as true elliptical galaxies are marked with a yellow star (this work) or a blue star (literature sample). The error bars represent the standard deviation of the Galaxy Zoo sample, or the measurement error for individual SNe. }
\label{fig:seds} 
\end{figure*}     

    A few cases of CCSNe in the outskirts of non-star forming elliptical galaxies have been reported. SN\,2016hil\ was a SN II found at a relatively large offset of $~26\ \rm kpc$ from the center of a massive elliptical galaxy \citep{Irani2019b}. The SN Ibn PS1-12sk was found at a similar offset from a massive elliptical galaxy \citep{sanders2013}. In both cases, the large offset from the massive nearby elliptical galaxy might indicate that the association is spurious, although deep observations of the SN site using the \textit{Hubble Space Telescope} (\textit{HST}) and Keck argue against an underlying faint host-galaxy \citep{Hosseinzadeh2019,Irani2019b}.

Hubble first classified galaxies using a "tuning fork" scheme \citep{Hubble1926}, using the presence of spiral and bar features. This Scheme was later improved (e.g., by \citealt{DeVaucouleurs1959, Sandage1961}) to include finer morphological features. Galaxy morphology is broadly categorized into bulge dominated, "red and dead", early-type galaxies, and blue star-forming late-type galaxies. While the colors, morphologies and star-formation of galaxies are tightly correlated, they do not map exactly onto each other \citep{Strateva2001, Trager2000, DeLucia2006, Buta2011}. The observed bimodality in galaxy colors and star-formation properties requires a physical mechanism to quench star-formation in massive evolved galaxies; a topic of ongoing research \citep[e.g.,][]{Man2018, Gabor2010}.

While generally considered passive galaxies, the low-level star-formation still ongoing in some early-type galaxies is a topic of ongoing research.  \cite{Kaviraj2008} found that up to $\sim 30 \%$ of early-type galaxies might have undergone recent episodes of star formation, leading to bluer UV-optical colors. \cite{hernandez2009} further suggest that the blue colors of early-type galaxies with $4<NUV - \textit{r}_{\rm PS1}<5.4$ mag could be explained by a recent low-level star-forming episode, or by the presence of extreme horizontal branch (EHB) stars. \cite{Petty2013} also suggested EHB stars are the source of the blue UV-optical colors of UV excess early-type galaxies.  \cite{Salim2012} studied a sample of early-type galaxies from the SDSS survey with a strong UV excess and found structured UV morphology in 93\% of their sample. In 75\% of the cases, the star formation extended to offsets of 25-75 kpc, indicative of galaxy scale inside-out growth, fueled by accretion of gas from the intergalactic medium (IGM). Such growth is found to occur for massive galaxies \citep[][]{SanchezBlazquez2007,perez2013}. The rest of their sample is characterized by patchy and centered (5-15 kpc) star formation. On the other hand, \cite{Gomes2016} found star-forming lanes in 3 nearby early-type galaxies, documenting the still ongoing star-formation in these galaxies. Assuming the ongoing star-formation in elliptical galaxies is similar to star-formation in spirals, a population of SNe is expected to explode in these environments. 

New transient surveys such as the Zwicky Transient Facility (ZTF; \citealt{Bellm2019b,Graham2019,Dekany2020}), the Astroid Terrestrial Last Alert System (ATLAS; \citealt{Tonry2018}), the Panoramic Survey Telescope and Rapid Response System 1 (PS1; \citealt{Chambers2016}), and the All Sky Automated Survey for Supernovae (ASAS-SN; \citealt{shappee2014}), monitor the entire nightsky with a high cadence and to unprecedented depths, and discover thousands of SNe in the process. This allows SNe to be used as probes to trace star-formation in elliptical galaxies where star-formation was thought to have ceased. The ZTF Bright Transient Survey (BTS)  \citep{fremling2020,perley2020} is the largest untargeted spectroscopically complete SN survey to date. It classifies 93\% of all SNe with peak magnitudes $m_{peak}<18.5\ \rm mag$ and 75\% of all SNe with $m_{peak}<19\ \rm mag$. This sample allows for a systematic study of residual CCSN populations in low-redshift environments. 

In this paper, we used the BTS to conduct a search for spectroscopically confirmed CCSNe in elliptical host galaxies. We present observations of three such SNe. We combine these events with a sample of literature objects, analyze the SN properties, and characterize their host environments. This paper is organized as follows: In Section \ref{sec:sample} we describe our sample selection and the comparison samples. In Section \ref{sec:observations} we present the spectroscopic and photometric observations of the SNe and their hosts. In Section \ref{sec:results} we characterize the transients and their host galaxies. We discuss the implications of these results and present our conclusions in Section \ref{sec:discussion}. Throughout this paper we assume $H_0=73~{\rm km\,s}^{-1}\,{\rm Mpc}^{-1}$ and a $\Lambda$CDM cosmology with $\Omega_m=0.27$ and $\Omega_\Lambda=0.73$ (Wilkinson Microwave Anisotropy Probe 3 year results; \citealt{Spergel2007}). All magnitudes are reported in the AB system and are corrected for line-of-sight reddening based on \citet{schlafly2011}; see Sec. \ref{extinction}.\\

\section{Sample}
\label{sec:sample}

\subsection{Candidate selection process}
\label{subsec:candidate}

As of December 30, 2020, the BTS sample contains 4018 spectroscopically classified SNe. Of these, we select SNe satisfying the following criteria:
\begin{enumerate}

\item The redshift is $0.015<z<0.1$. The lower bound is meant to avoid shredding of nearby galaxies into multiple sources in galaxy catalogs and the upper bound is due to the reduced completeness of galaxy redshift catalogs \citep{fremling2020}, required to ensure the SN and the galaxy are indeed associated. 

\item The likely host galaxies of the ZTF BTS SNe are continuously identified in the BTS sample explorer\footnote{https://sites.astro.caltech.edu/ztf/bts/explorer.php} \citep{perley2020}. This is done by an automatic cross-matching of the SN positions with the nearest galaxies in the Pan-STARRS (PS1; \citealt{Chambers2016}) or SDSS   \citep{Alam2015} photometric catalogs. Host-galaxy matching is complete out to an offset of $<90 \arcsec$ and $<30$ kpc (projected distance) and employs a criterion to distinguish between multiple host-galaxy associations. We only select SNe which are associated with a host galaxy in this way.  
\end{enumerate}
Of the 4018 SNe in the sample 3855 had a redshift in the required range. Among those, we could identify the most likely host galaxy of 3330 SNe. We cross-match these host-galaxies with the \textit{Galaxy Evolution Explorer} (\galex) Data Release (DR) 8/9 \citep{Martin2005a} and the ALLWISE \citep{Cutri2014} catalogs and obtain the FUV (1542\AA), NUV (2274\AA) and $W1$-$W4$ [33,500\AA - 220,000\AA] IR photometry. We correct the results for Galactic reddening using the maps of \cite{schlafly2011}.  Queries were performed from the VizieR Catalog \citep{Ochsenbein2000} using \package{astroquery} \citep{Ginsburg2019}. We use the host photometry to define a sample of CCSNe with early-type hosts.  These are here defined as:\\

\begin{enumerate}
    \item Galaxies with $W2-W3<0.5\ \rm mag$ and $NUV-\textit{r}_{\rm PS1}>3\ \rm mag$, if both the $W2-W3$ color and the $NUV - \textit{r}_{\rm PS1}$ colors are available.    
    \item Or  galaxies with $W2-W3<0.3$, if only the $W2-W3$ color is available.
\end{enumerate}

This definition is satisfied by 75 \% of elliptical galaxies in the Galaxy Zoo \citep{Lintott2011} sample. It is chosen so that it includes all regions in the color-color parameter space which have a higher number of ellipticals compared to spirals, calculated in 0.5 mag bins.

Next, we search for any SN not classified as a SN Ia, and examine the host morphology visually. To be considered a CCSN in an elliptical host, we require:

\begin{enumerate}
\item Independent and consistent spectroscopic redshift measurements of both the SN and the host galaxy from publicly available catalogs. 

\item Visual confirmation of the red color, and absence of a bar, spiral arms, or a disc structure in the deepest available images. 

\item Once aperture-matched photometry and source deblending is performed, the host galaxy still occupies our double-color region for elliptical galaxies.

\item Confirmation of a CCSN classification: SN classifications in the BTS \citep{fremling2020} are made using both human inspection and using \package{SuperNova IDentification} (SNID; \citealt{Blondin2007}), and are thus generally reliable. However since CCSNe generally avoid elliptical galaxies, we take special care to avoid misclassified SNe Ia and require verification of the SN type through a visual inspection of the spectrum and by fitting the spectra using a python adaptation of \package{Superfit} \citep{Howell2005} with an updated spectrum template bank (Goldwasser et al., in prep).
\end{enumerate}

21 CCSNe pass our color criteria, of which only four SN hosts emerged as having elliptical morphology - \uik\ (SN II), \fsh\ (SN II), \ape\ (SN Ic) and SN\,2019cmv (SLSN-II). While SN\,2019cmv passes our sample criteria, we exclude it due to the combination of having a high-offset from the nearby elliptical galaxy, and the shallow limits on an underlying host at the SN site. At the distance of SN\,2019cmv we cannot rule out the presence of a faint underlying host with a brightness of $>-14\ \rm mag$. Such low-mass hosts have been observed for several SLSNe \citep{perley2016,Schulze2018}, and are the most likely explanation for the unusual location of SN\,2019cmv. SN\,2020oce, classified as a SN Ic in the BTS, also passes our sample criteria, but we find that it is fit well by spectra of 91bg-like SNe Ia, and so exclude it from our sample.
In Fig.~\ref{fig:all_hosts} we show PS1 image cutouts of the three CCSN host galaxies from the BTS matching our criteria. The images are constructed using the method described in \cite{Lupton2004}.\\

\indent The data points in the left panel of Fig. \ref{fig:color_color} show the UV-optical and mid-infrared (MIR) colors of CCSN hosts compared to the colors of elliptical and spiral galaxies from the Galaxy Zoo catalog \citep{Lintott2011}. In this plot, a good separation is achieved between the colors of elliptical and spiral galaxies. On the right panel of Fig.~\ref{fig:color_color} we show the same color-color plot for the host galaxies of SNe Ia. As expected, the vast majority of CCSNe occur in the region occupied by spirals. In Fig. \ref{fig:seds} we show the mean observed colors of spirals and ellipticals from the Galaxy Zoo catalog normalized to the $W2$ band, and compare them with the colors of elliptical host galaxies of CCSNe. As this figure shows, the most significant difference between the spectral energy distribution (SED) of spiral and elliptical galaxies is seen in the UV and in the MIR. The elliptical host galaxies of CCSNe have SEDs similar to galaxy zoo ellitpicals.

\subsection{Literature sample}
\label{subsec:literature}
In addition to the objects presented in this paper, we compile a sample of CCSNe in elliptical galaxies from the literature. We apply the criteria in Sec. \ref{subsec:candidate} to the CCSN sample \citep{Schulze2020} from the Palomar Transient Factory (PTF; \citealt{law2009}) and the intermediate Palomar Transient Factory (iPTF; \citealt{kulkarni2013}), and to published CCSNe from \cite{suh2011}, \cite{Hakobyan2008}, \cite{Graham2012} and \cite{sanders2013}. Our findings generally agree with the conclusions of these papers - that most CCSNe near elliptical galaxies occurred in either misclassified spirals or star-forming ellipticals. A small minority of the SNe analyzed in these papers do pass our criteria, and we combine these with our BTS objects to form a combined sample. We further require that objects in our sample have public spectra, so that the classification can be confirmed. In total, our combined sample includes 10 SNe, listed in Table~\ref{tab:lit_objs}, along with their classifications and estimated peak luminosities. We note that the host of the Type II SN Abell~399 11 19 0 \citep{Graham2012} did not pass our sample inclusion criteria, since it is not detected in both the $NUV$ and $W3$ bands. However, the available limits ($W2-W3 < 0.06\ \rm mag$, $NUV - \textit{r}_{\rm PS1}>4.5$ mag), along with the galaxy morphology, indicate this host is also an elliptical galaxy.

\begin{deluxetable*}{lccccclll}
\label{tab:lit_objs}
\centering
\tablecaption{Sample of CCSNe with elliptical hosts analyzed in this work.}

\tablewidth{20pt} 

\tablehead{\colhead{SN} & \colhead{R.A. (J2000)} & \colhead{Dec. (J2000)} & \colhead{Redshift} & \colhead{Projected offset (kpc)} &\colhead{Peak $M_{\textit{r}}$ (AB mag)$^{a,b}$} & \colhead{SN type} & \colhead{Reference}} 
\tabletypesize{\scriptsize} 
\startdata
SN2003ky & 179.52625 & 47.33319 & 0.047 & 8.3 & $-19.3$  & II  & \cite{Armstrong2003}\\
SN2006ee & 29.8975 & 14.00544 & 0.015 & 4.3 & $-16.64$    & II  & \cite{Puckett2006}\\
SN2006gy & 49.36275 & 41.40542 & 0.019 & 1.1 & $-20.2$   & IIn & \cite{ofek2007,Smith2007}\\
PTF10gqf & 225.96897 & 55.62717 & 0.045 & 29.0 & $-16.5$     & II  & \cite{Schulze2020}\\
PS1-12sk & 131.22858 & 42.97136 & 0.054 & 32.9 & $-19.2$     & Ibn & \cite{sanders2013}\\
SN2016hil & 17.603106 & 14.204318 & 0.061 & 26.4 & $-17.0$   & II  & \cite{Irani2019b}\\
PTF16pq & 139.39429 & 18.95197 & 0.028 & 4.0 & $-16.1$         & II  & \cite{Schulze2020}\\
SN2018fsh & 127.73595 & 39.83588 & 0.029 & 22.8 & $-18.2$& II  & This work\\
SN2019ape & 162.92728 & 18.4813 & 0.020 & 4.8 & $-16.83$     & Ic  & This work\\
SN2020uik & 120.47579 & -6.76099 & 0.028 & 8.6 & $-17.22$ & II  & This work\\
\enddata
\tablenotetext{a}{All measurements are taken from the Open Supernova Catalog \citep{Guillochon2017} or from cited references.}
\tablenotetext{b}{Observed \textit{r}-band magnitudes were adopted when available. Otherwise, an average between R and V bands was adopted. If no color information was available, we adopt the peak magnitude in the observed band. Magnitudes were corrected for Galactic reddening.} 
\vspace{-0.15cm}
\end{deluxetable*}

\subsection{Comparison samples}
\label{subsec:comparison}
Throughout this work, we use several SN and galaxy
samples for comparison. These are:

\begin{itemize}
    \item BTS CCSNe: we selected all BTS SNe classified as CCSNe that occured in host galaxies with measured \textit{WISE} $W2-W3$ and UV-optical colors. This sample consists of 478 objects, of which 465 fall outside of the region that we associate with elliptical galaxies in Fig. \ref{fig:color_color}.  
    \item SNe Ia elliptical galaxies: we selected all BTS SNe \citep{perley2020} classified as SNe Ia that occured in host galaxies with both $W2-W3<0.5\ \rm mag$ and $NUV-\textit{r}_{\rm PS1}>3\ \rm mag$. This sample contains 240 objects. 
    \item Galaxy Zoo \citep{Lintott2011}  galaxies: we randomly selected 50,000 galaxies from the Galaxy Zoo sample that have morphological classifications and with a redshift in the range $0.015<z<0.05$. Of these galaxies, 1309 (10265) were classified as elliptical (spiral). Queries were performed using \package{CasJobs} \citep{OMullane2005}. We note that \cite{Lintott2011} use a broad categorization of galaxies into ellitpicals and spirals. The elliptical group includes both elliptical (E) and lenticular (S0) galaxies.

\end{itemize}

\subsection{SN discovery and classification}
\label{subsec:detection & classification}
\subsubsection{ \fsh }
\fsh\ was detected in the ZTF alert stream on UT 2018 Aug. 31.52 (JD 2458362.02) at J2000 coordinates of  $\alpha=08^{h}30^{m}56.6^{s}$, $\delta=+39^{\circ}50^{'}09.2^{"}$ at a brightness of $r=19.0$ mag \citep{2018TNSTR1295....1F}.
Based on a spectrum obtained on UT 2018 Sep. 17.56 (JD 2458379.06) the transient was classified as a SN II \citep{fremling_tns_2018} at a redshift of $z=0.029$. This redshift is consistent with the value ($z=0.029084$) listed in the NASA Extragalactic Database (NED)\footnote{https://ned.ipac.caltech.edu/} for MCG +07-18-013, a red galaxy offset by  $38.6 \arcsec$ from the SN location. The host redshift corresponds to a distance of $122\pm 9\ \rm Mpc$ corrected for Virgo, Shapley and Great attractor infall \citep[via NED]{Mould2000}. ZTF did not observe the field in the three months before the first SN detection and the last ZTF non-detection is on 2018 May 25.17. The SN is visible in data by ATLAS as early as UT 2018 Jun 15.26, $\sim84$ days before the first ZTF detection and 4 days after the previous non-detection. We adopt UT 2018 Jun 13.25 (JD $2458282.75$), the midpoint between the last non-detection and the first detection, as an estimate of the explosion time for this event.   
\begin{table*}
\centering
\caption{Photometric observations of SN\,2018fsh, SN\,2019ape and SN\,2020uik.}
\label{tab:phot}

\begin{tabular}{lccllc}
\toprule
SN name & JD & Estimated time from explosion (rest-frame days)  & Instrument & Filter	    & AB Magnitude\\	
\midrule
\ape & 2458514.94 & 1.47 & ZTF & \textit{r} & $20.45 \pm 0.19$ \\
\ape & 2458522.92 & 9.29 & ZTF & \textit{g} & $18.72 \pm 0.01$ \\
\ape & 2458524.45 & 10.79 & LT & \textit{g} & $18.63 \pm 0.01$ \\
\ape & 2458524.45 & 10.79 & LT & \textit{r} & $18.07 \pm 0.01$ \\
\ape & 2458524.45 & 10.79 & LT & \textit{z} & $18.14 \pm 0.02$ \\
\ape & 2458524.45 & 10.79 & LT & \textit{i} & $18.19 \pm 0.01$ \\
\ape & 2458526.93 & 13.22 & ZTF & \textit{g} & $18.67 \pm 0.03$ \\
\ape & 2458528.0 & 14.27 & LT & \textit{g} & $18.63 \pm 0.01$ \\
\fsh & 2458284.76 & 4.63 & ATLAS & c & $17.14 \pm 0.03$ \\
\fsh & 2458286.76 & 6.57 & ATLAS & o & $17.22 \pm 0.09$ \\
\fsh & 2458371.01 & 88.44 & ZTF & \textit{r} & $18.78 \pm 0.02$ \\
\fsh & 2458373.99 & 91.34 & ZTF & \textit{r} & $18.78 \pm 0.02$ \\
\fsh & 2458374.01 & 91.36 & ZTF & \textit{g} & $20.21 \pm 0.11$ \\
\fsh & 2458376.99 & 94.26 & ZTF & \textit{g} & $20.3 \pm 0.05$ \\
\fsh & 2458377.02 & 94.28 & ZTF & \textit{r} & $18.86 \pm 0.02$ \\
\uik & 2459109.11 & 23.41 & ATLAS & o & $18.24 \pm 0.07$ \\
\uik & 2459111.1 & 25.34 & ATLAS & c & $18.43 \pm 0.04$ \\
\uik & 2459113.12 & 27.3 & ATLAS & o & $18.27 \pm 0.03$ \\
\uik & 2459115.09 & 29.21 & ATLAS & c & $18.53 \pm 0.07$ \\
\uik & 2459115.1 & 29.22 & ATLAS & c & $18.33 \pm 0.12$ \\
\uik & 2459126.12 & 39.92 & ATLAS & o & $18.11 \pm 0.02$ \\
\uik & 2459127.12 & 40.89 & ATLAS & o & $18.01 \pm 0.1$ \\
\uik & 2459130.97 & 44.63 & ZTF & \textit{r} & $18.23 \pm 0.1$ \\

\bottomrule     
\end{tabular}
\tablecomments{All measurements are reported in the AB system and are corrected for Galactic line of sight reddening.\\
The full table will be made available electronically on the journal website and on WISeREP upon publication.}
\end{table*}

\subsubsection{\ape}
\label{sec:det_ape}
\ape\ was detected in Pan-STARRS1 data on UT 2019 Feb. 05.39 (JD 2458519.89) at J2000 coordinates of  $\alpha=10^{h}51^{m}42.5^{s}$, $\delta=+18^{\circ}28^{'}52.73^{"}$ at a brightness of $w=18.57$ mag \citep{Chambers2019_tns}.
On UT 2019 Feb. 08.71 (JD 2458523.21) the transient was classified as a SN Ic \citep{Carini2019} at a redshift of $z=0.02$ based on the SN features. \citealt{De2020} acquired a nebular spectrum of the SN  $300\ \rm days$ after explosion and conclude it does not belong to the Ca-rich group based on the \CaII ~to [O I] ratio. \ape\ was detected in the massive elliptical NGC 3426 ($z=0.020414$). We adopt the host redshift and a corresponding redshift-dependent distance of $90.4\pm6.4\ \rm Mpc$ corrected for Virgo, Shapley and Great attractor infall. Applying the ZTF forced photometry pipeline \citep{Masci2019} yields an earlier detection from UT 2019 Jan. 31.44 (JD 2458514.94), at a brightness of $r=20.52\pm0.19\ \rm mag$. The latest non-detection is at $>20.9$ mag three days earlier.  We estimate the explosion date as JD $ 2458513.44\pm{1.5}$, the midpoint between the first detection and the last non-detection.
\subsubsection{\uik}
\uik\ was detected in ATLAS data on UT 2020 Sep. 20.62 (JD 2459113.12) at J2000 coordinates of  $\alpha=08^{h}01^{m}54.18^{s}$,$\delta=-06^{\circ}45^{'}39.52^{"}$ at a brightness of $c=18.5$ mag \citep{Tonry2020_tns}.
On UT 2020 Oct. 18.53 (JD 2459141.03) the transient was classified as a SN II \citep{Dahiwale2020} at a redshift of $z=0.03$ based on the SN features. These are consistent with the NED redshift of $z=0.028156$ of WISEA J080154.84-064527.1, a red galaxy offset by $15.0  \arcsec$ from the SN location. We adopt the host redshift and a corresponding redshift-dependent distance of $118.1\pm8.3\ \rm Mpc$ corrected for Virgo, Shapley and Great attractor infall. \uik\ was detected during a plateau in its light curve after a long period that the field was not observed. Assuming a plateau that extends $~100$ days since explosion \citep{arcavi2017}, we estimate an explosion time of around JD $~2459085$ - 28 days prior to the first detection.  The best fitting \package{Superfit} template of its first spectrum, taken 56 days after the estimated explosion, is the spectrum of the SN II SN\,2013fs (taken 57 days after explosion; \citealt{Yaron2017}).

\subsection{Extinction}
\label{extinction}
 
We queried the NASA/IPAC NED Galactic 
Extinction Calculator\footnote{https://ned.ipac.caltech.edu/forms/calculator.html} for the foreground Galactic extinction in the line-of-sight toward each of the 3 SNe presented in this work, derived from the maps of \cite{schlafly2011}. We find a line-of-sight extinction of E$_{\rm (B-V, MW)}=0.043\ \rm mag$ for \fsh\, E$_{\rm (B-V, MW)}=0.031\ \rm mag$ for \ape\, and E$_{\rm (B-V, MW)}=0.082\ \rm mag$ for \uik.  
We estimate the host extinction of \ape\ (along the line-of-sight to the SN) using the $\textit{g-r}$ color curve of \ape\ compared to intrinsic color curves of SNe Ic (\citealt{stritzinger2018b}, \citealt{Drout2011}). We find E$_{\rm (B-V, Host)}=0.14\pm0.03\ \rm mag$. This value is broadly consistent with the value derived using the equivalent width of the Na D doublet absorption feature \citep{poznanski2012}. We do not attempt to estimate the host-extinction for \fsh\ or \uik, but estimate these are not significant due to the large offset from their host galaxies. 

\section{Observations}
\label{sec:observations}

\subsection{Photometry}
\label{subsec:photometry}
For all SNe discussed in this paper, we acquired \textit{gri} photometry using the P48 ZTF camera \citep{Dekany2020}. These data were processed using the ZTF data processing system \citep{Masci2019}. Light curves were obtained using the forced photometry pipeline \citep{Masci2019} on difference images produced using the optimal image subtraction algorithm of \citet{Zackay2016} at the position of the SN, as reported by the first ZTF alert. We report detections above a $3\sigma$ threshold. Same-night detections were binned in order to boost the signal. Additional photometry was acquired with:
\begin{itemize} 

\item The Las Cumbres Observatory (LCO) network of 1m telescopes through the Global Supernova Project \citep{Howell2019}. Photometric data were reduced using the \package{lcogtsnpipe} pipeline which performs PSF-fitting photometry. Landolt standard field stars were used to calculate the zeropoints for the filters $UBV$; whereas for \textit{gri} bands we use Sloan magnitudes of stars in the same field as the object.

\item The two ATLAS 0.5m telescopes on
Haleakala and Mauna Loa, Hawaii, USA \citep{Tonry2018}. Data were reduced using the forced photometry service \citep{Smith2020}. 

\item The 2.0 m robotic Liverpool Telescope (LT; \citealt{Steele2004}) at the Observatorio del Roque de los Muchachos Observatory on La Palma using the  optical imager (IO:O) through the $g$, $r$, $i$, and $z$ bands. Photometry was reduced using standard {\sc iraf} routines within a custom {\sc python} script and stacked using \package{SWarp} \citep{Bertin2002}. Digital image subtraction was performed versus PS1 reference imaging following the techniques of \citet{fremling2016} and calibration was performed relative to PS1 photometric standards.

\item The Rainbow Camera \citep{Blagorodnova2018a} on the Palomar 60-inch telescope (P60; \citealt{Cenko2006}). Reductions were performed using the automatic pipeline described by \cite{fremling2016}. 
\end{itemize}

\begin{figure*}
\begin{centering}
\includegraphics[width=1.6\columnwidth]{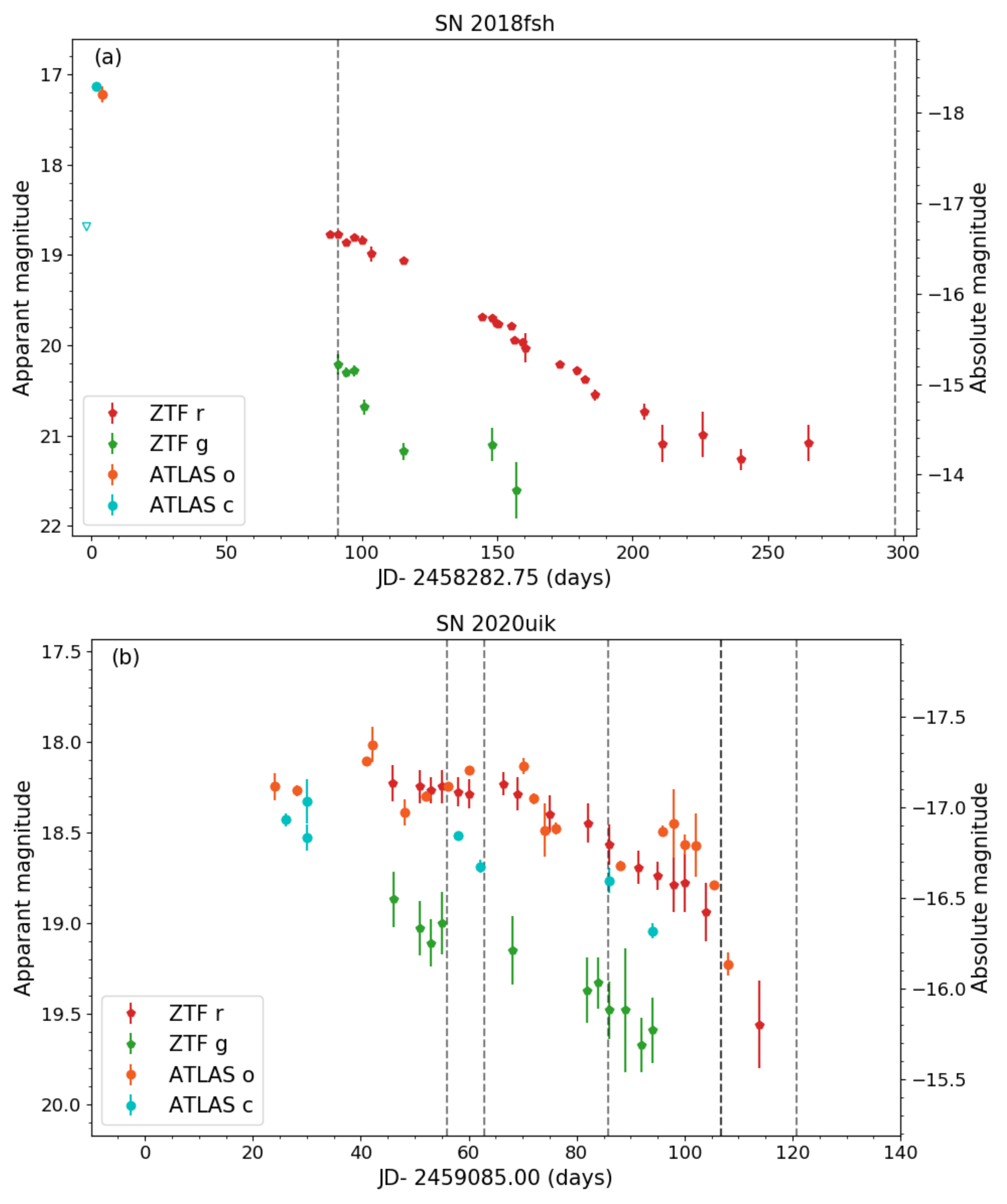}
\caption{Light curves of SN\,2018fsh (a) and SN\,2020uik (b).  Full symbols are detections and empty triangles are $5\sigma$ upper limits. Vertical black dashed lines mark the dates of spectroscopic observations.}  \label{fig:18fsh_20uik_lc}
\end{centering}

\end{figure*}

\begin{figure*}
\centering
\includegraphics[width=2\columnwidth]{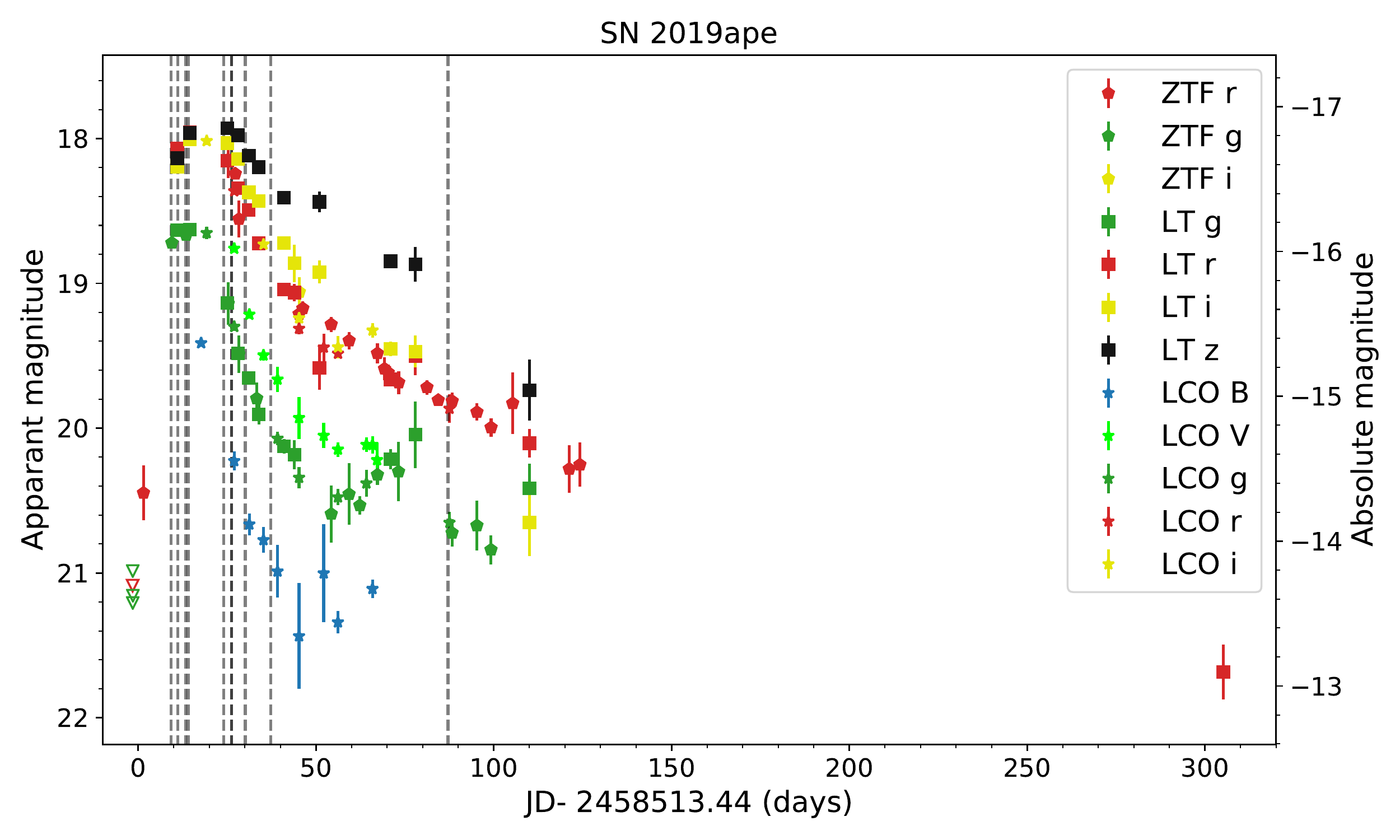}
\caption{Optical light curves of SN2019ape. Full symbols are detections and empty triangles are P48 $5\sigma$ upper limits. Vertical black dashed lines mark the dates of spectroscopic observations. }  
\label{fig:19ape_lc}
\end{figure*}

For \ape, instrument cross-calibration was performed by applying constant shifts calculated using polynomial fits to band-specific light curves. In case a single instrument was available for a given band, the calibration was performed using synthetic photometry on the spectra scaled to \textit{r}-band photometry to ensure accuracy relative to other bands. Following these procedures a constant shift of $-0.32\ \rm mag$ was applied to LCO \textit{g}-band light curves to match the evolution of LT and ZTF light curves. No other offsets were required. The photometry in this work will be made available through the Weizmann Interactive Supernova data REPository (WISeREP;\citealt{yaron2012}) upon publication.  
The light curves of \fsh\ and \uik\ are shown in Fig. \ref{fig:18fsh_20uik_lc}, and the light curves of \ape\ are shown in Fig. \ref{fig:19ape_lc}. 

\subsection{Spectroscopy}
\label{subsec:spectroscopy}

Spectroscopic follow-up of SNe appearing in this work was performed using a variety of instruments:
\begin{itemize} 

\item The 3.6 m ESO New Technology Telescope (NTT) at La Silla, Chile, using the ESO Faint Object Spectrograph and Camera (v.2) (EFOSC2; \citealt{Buzzoni1984}) as part of ePESSTO+. These data were reduced using the PESSTO pipeline \citep{smartt2015b}. 

\item The Nordic Optical telescope (NOT) using The Alhambra Faint Object Spectrograph and Camera (ALFOSC). Reductions were performed using {\sc foscgui}\footnote{\url{http://graspa.oapd.inaf.it/foscgui.html}}.

\item The 200-inch Hale telescope at Palomar observatory using the Double Beam Spectrograph \citep{Oke1982}. These data were reduced following standard procedures using the P200/DBSP pipeline described in \citet{Bellm2016}.

\item The Spectral Energy Distribution machine (SEDm; \citealt{Blagorodnova2018a}) mounted on the Palomar 60-inch telescope. Data were reduced using the automatic SEDm pipeline \citep{rigault2019}. 

\item The Spectrograph for the Rapid Acquisition of Transients (SPRAT; \citealt{Piascik2014}) on the Liverpool Telescope. SPRAT spectra were reduced using the LT pipeline \citep{Barnsley2016} and flux calibrated using a custom {\sc python} routine.

\item The Low Resolution Imaging Spectrometer (LRIS; \citealt{Oke1995}) on the 10-m Keck I telescope. The data were reduced using the  \package{LRIS automated reduction pipeline} (\package{LPipe}; \citealt{Perley2019}). 
\end{itemize}

\begin{figure*}
\centering

\includegraphics[width=1.8\columnwidth]{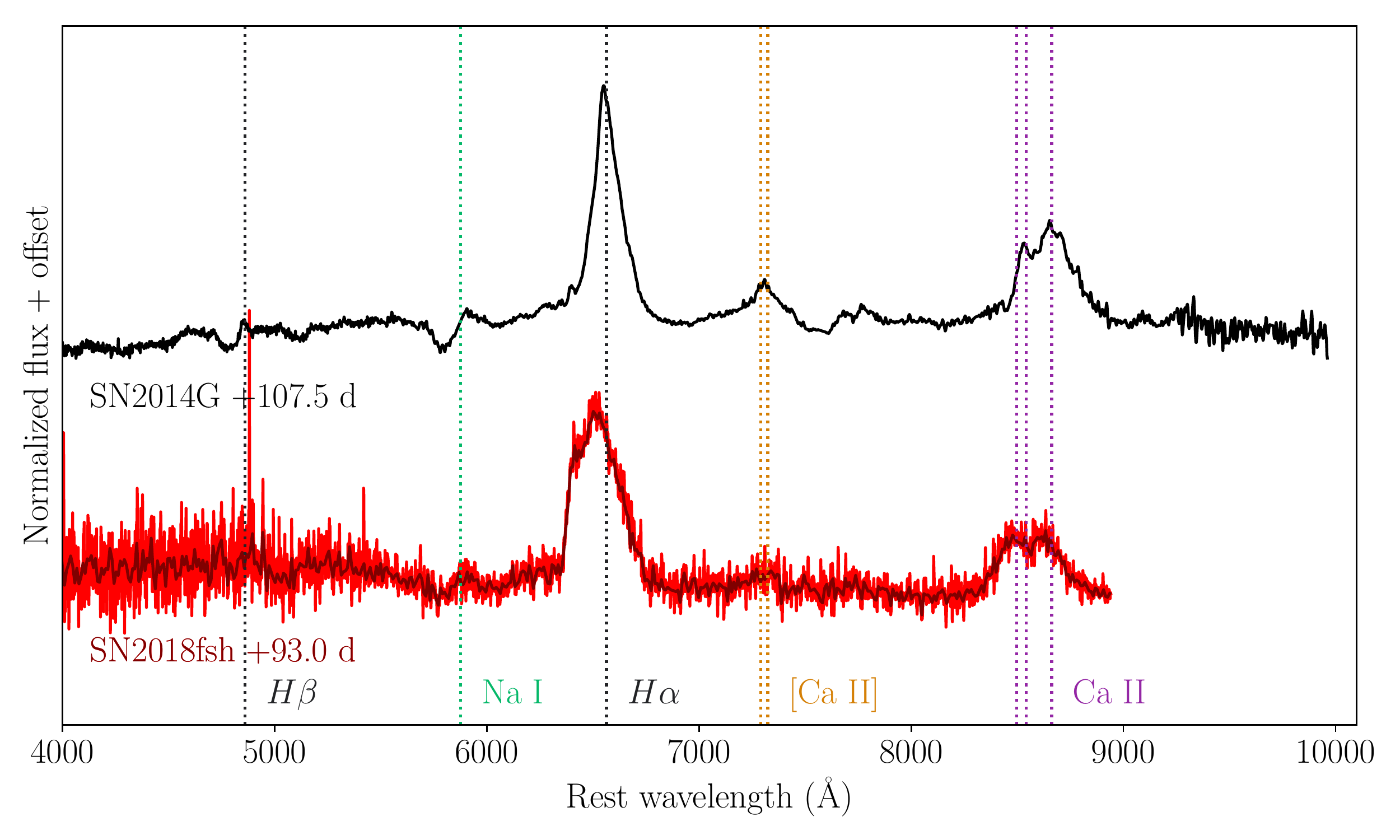}
\caption{Comparison of the spectrum of \fsh\ to the spectrum of the best matching \package{Superfit} result SN\,2014G at a similar phase. The main elements appearing in both spectra are marked with vertical lines. The dark line is a binned version of the spectrum to guide the eye.} 
\label{fig:18fsh_spec}

\end{figure*}

\begin{figure*}
\centering
\includegraphics[width=1.6\columnwidth]{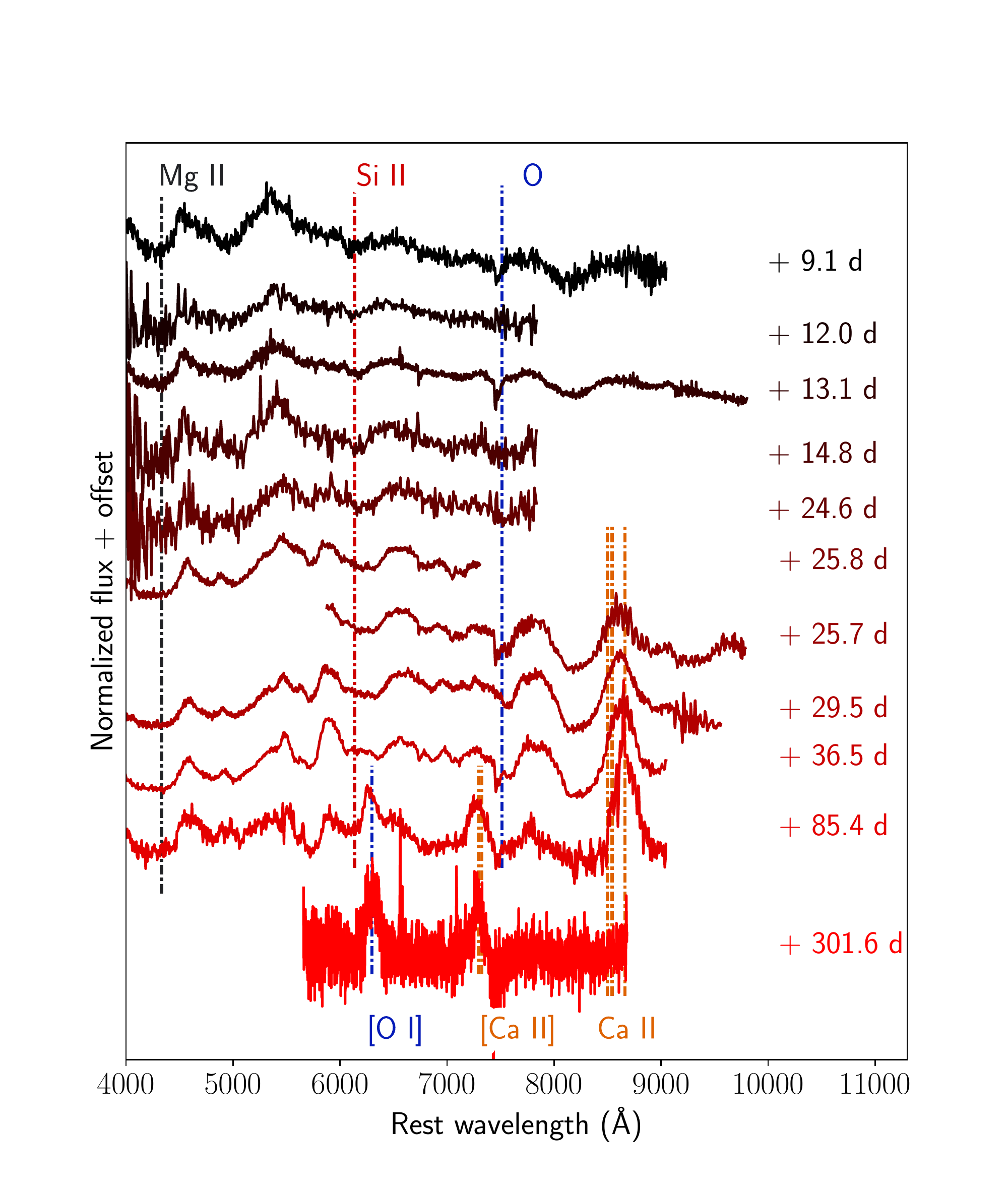}
\caption{Spectral sequence of SN\,2019ape. Phase is denoted in rest frame days. The main features appearing in the spectrum are marked with vertical lines. Absorption features are blue shifted by $10,000\ \rm km\ s^-{-1}$ } 
\label{fig:19ape_spec}

\end{figure*}

\begin{figure*}
\centering

\includegraphics[width=1.3\columnwidth]{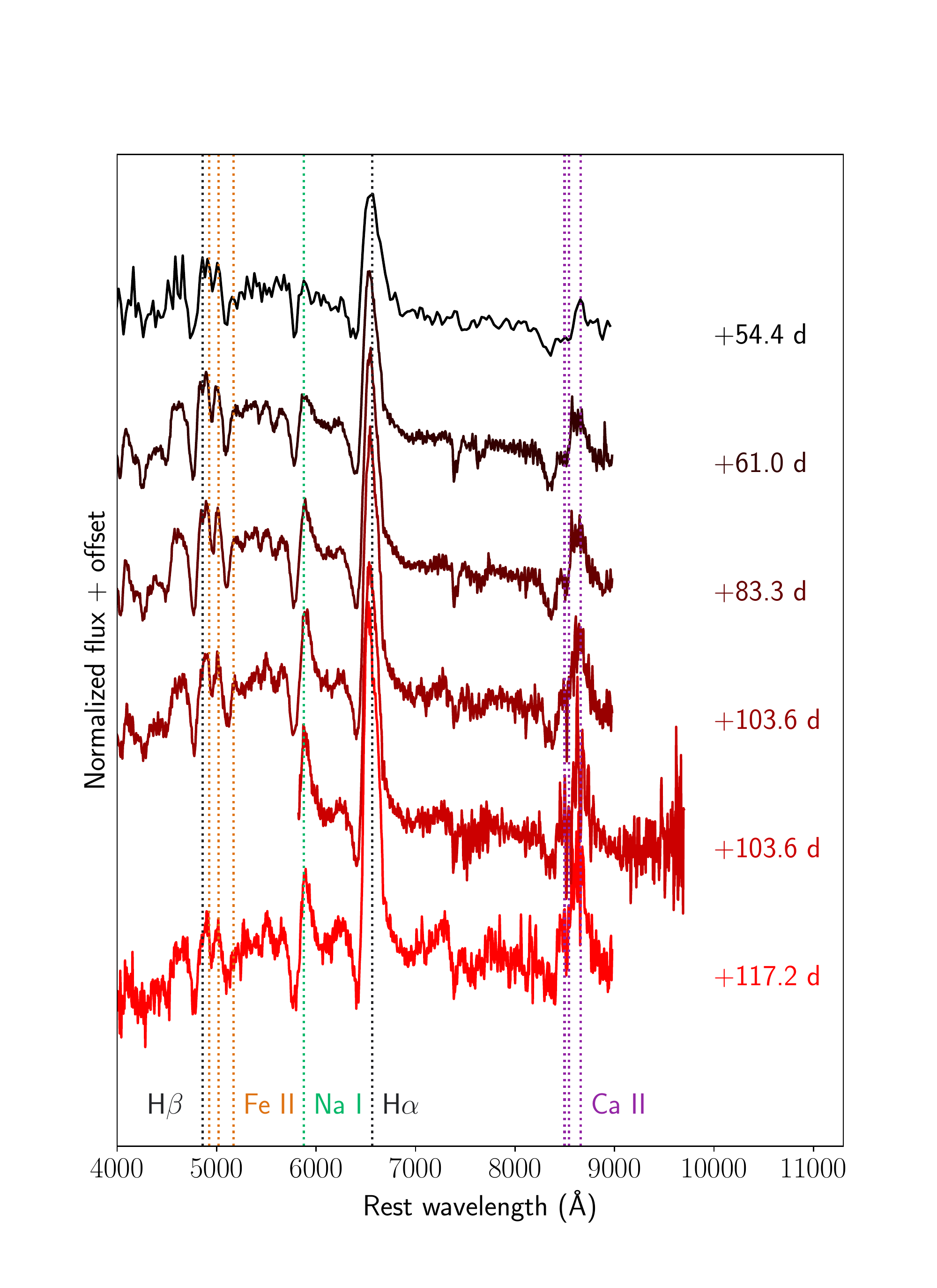}
\caption{Spectral sequence of SN\,2020uik.  Phase is in rest frame days.} 
\label{fig:20uik_spec}

\end{figure*}

Figure \ref{fig:18fsh_spec} shows the spectrum of \fsh\ 96 days after explosion in comparison with the best fitting SN II spectrum at a similar phase using \package{Superfit}. Figures \ref{fig:19ape_spec} and \ref{fig:20uik_spec} show the spectral evolution of \ape\ and \uik\ respectively. Table \ref{tab:opt-spec} contains a log of spectroscopic observations presented in this work. 
All spectra will be made available to the public on WISeREP\footnote{https://www.wiserep.org/} upon publication.

\begin{deluxetable*}{llclccc} 
\tablecaption{Log of spectroscopic observations}
\label{tab:opt-spec}

\tablewidth{20pt} 

\tablehead{\colhead{SN} & \colhead{Date} & \colhead{Estimated time since explosion (rest-frame days) } & \colhead{Instrument} & \colhead{Exposure time (s)} & \colhead{Airmass} &\colhead{grism}}
\tabletypesize{\scriptsize} 

\startdata 
SN\,2019ape & 2019 Feb. 08 & 9.75 & NTT/EFOSC2  & 900           & 1.49    & 13       \\
SN\,2019ape & 2019 Feb. 10 & 12.56 & LT/SPRAT   & 1200          & 1.02    & red      \\
SN\,2019ape & 2019 Feb. 12 & 14.33 & P200/DBSP  & 900           & 1.05    & 600/4000 \\
SN\,2019ape & 2019 Feb. 13 & 15.39 & LT/SPRAT   & 1200          & 1.19    & red      \\
SN\,2019ape & 2019 Feb. 23 & 25.29 & LT/SPRAT   & 1200          & 1.05    & red      \\
SN\,2019ape & 2019 Feb. 25 & 26.32 & NTT/EFOSC2 & 2700          & 1.59    & 16       \\
SN\,2019ape & 2019 Feb. 25 & 26.37 & NTT/EFOSC2 & 2700          & 1.49    & 11       \\
SN\,2019ape & 2019 Mar. 01 & 30.15 & NOT/ALFOSC & 2400          & 1.02    & 4        \\
SN\,2019ape & 2019 Mar. 08 & 37.12 & NTT/EFOSC2 & 2700          & 1.48    & 13       \\
SN\,2019ape & 2019 Apr. 27 & 86 & NTT/EFOSC2    & $2\times2700$& 1.51    & 13       \\
SN\,2019ape & 2019 Dec. 03 & 302.19 & Keck/LRIS & 1750          & 1.04    & 600/4000,400/8500  \\
SN\,2020uik & 2020 Oct. 18 & 39.83  & P60/SEDM    & 2250           & 1.52     & IFU     \\
SN\,2020uik & 2020 Oct. 25 & 46.44 & NTT/EFOSC2   & 2700          & 1.13    & 13       \\
SN\,2020uik & 2020 Nov. 17 & 68.69 & NTT/EFOSC2  & 2700           & 1.35     & 13       \\
SN\,2020uik & 2020 Dec. 08 & 89.03 & NTT/EFOSC2   & 2700          & 1.35     & 13      \\
SN\,2020uik & 2020 Dec. 08 & 89.06& NTT/EFOSC2   & 2700          & 1.19     & 16      \\
SN\,2020uik & 2020 Dec. 22 & 102.6 & NTT/EFOSC2   & 2700          & 1.27     & 13       \\
SN\,2018fsh & 2018 Sep. 12 & 91.2 & P200/DBSP   & $2 \times 600$   & 1.66     & $316/7500,600/4000$ \\
SN\,2018fsh & 2019 Apr. 06 & 291.1 & Keck/LRIS   & $1750$   & 1.655     & $316/7500,600/4000$ \\
\enddata 

\end{deluxetable*}

\subsection{Host-galaxy photometry}
\label{subsec:host_photometry}
 We retrieved archival images of all SN host galaxies discussed in this work from \textit{Galaxy Evolution Explorer} (\galex) Data Release (DR) 8/9 \citep{Martin2005a}, SDSS DR9 \citep{Ahn2012a}, PS1 DR1 \citep{Chambers2016}, the Two-Micron All Sky Survey \citep[2MASS;][]{Skrutskie2006a}, and the unWISE \citep{Lang2014a} images from the \textit{NEOWISE} \citep{Meisner2017a} Reactivation Year 3. For SNe included in our sample we use the matched-aperture photometry software package \package{Lambda Adaptive Multi-Band Deblending Algorithm in R} \citep[\package{LAMBDAR};][]{Wright2016a} that is based on a photometry software package developed by \citet{Bourne2012a} and the tools presented by \cite{Schulze2020}. The photometry was either calibrated against zeropoints (\galex, PS1, SDSS, and \textit{NEOWISE}) or against a set of stars (2MASS). We correct the measurements for Milky-Way extinction based on \cite{schlafly2011}. The resulting photometry is summarized in Table \ref{tab:host_phot}. We note that in the case of SN\,2016hil, the $W3$ measurements are estimated from the $W2-W3$ color reported in \cite{Irani2019b}, where MIR photometry is measured from ALLWISE data using the same methodology described here.\\
 \\

\begin{deluxetable*}{ccccccc}[h]
\label{tab:host_phot}
\centering
\tablecaption{Photometry of SN host-galaxies analyzed in this work}
\tablewidth{26pt} 
\tablehead{\colhead{Instrument/Filter} & \colhead{$\lambda_{eff}$ ($\AA$)} & \colhead{SN\,2003ky} & \colhead{SN\,2006ee} & \colhead{SN\,2006gy} & \colhead{PTF10gqf} & \colhead{PS1-12sk}}
\tabletypesize{\scriptsize} 
\startdata 
GALEX/FUV & $1542$ & $18.92\pm 0.17$ & $17.83\pm 0.03$ & $20.25\pm 0.82$ & $19.83\pm  0.05$ & $19.27\pm 0.14$ \\
GALEX/NUV & $2274$ & $18.3\pm 0.08$ & $17.85\pm 0.06$ & $18.34\pm 0.24$ & $19.87\pm  0.06$ & $18.24\pm 0.06$ \\
SDSS/\textit{u} & $3595$ & $16.89\pm 0.04$ & $15.12\pm 0.05$ & $15.14\pm 0.14$ & $17.72\pm  0.09$ & $16.54\pm 0.05$ \\
SDSS/\textit{g} & $4640$ & $15.35\pm 0.01$ & $13.43\pm 0.01$ & $13.43\pm 0.08$ & $16.16\pm  0.02$ & $14.82\pm 0.04$ \\
SDSS/\textit{r} & $6122$ & $14.62\pm 0.01$ & $12.63\pm 0.01$ & $12.63\pm 0.04$ & $15.48\pm  0.01$ & $13.99\pm 0.01$ \\
SDSS/\textit{i} & $7440$ & $14.24\pm 0.01$ & $12.19\pm 0.01$ & $12.22\pm 0.04$ & $15.08\pm  0.01$ & $13.56\pm 0.01$ \\
SDSS/\textit{z} & $8897$ & $13.95\pm 0.02$ & $11.83\pm 0.03$ & $11.92\pm 0.03$ & $14.8\pm  0.04$ & $13.27\pm 0.04$ \\
$\rm PS1/\textit{g}$ & $4776$ & $15.38\pm 0.01$ & $13.39\pm 0.03$ & $13.4\pm 0.04$ & - & $14.79\pm 0.06$ \\
$\rm PS1/\textit{r}$ & $6130$ & $14.61\pm 0.01$ & $12.68\pm 0.02$ & $12.67\pm 0.09$ & $15.5\pm  0.01$ & $14.02\pm 0.06$ \\
$\rm PS1/\textit{i}$ & $7485$ & $14.24\pm 0.01$ & $12.28\pm 0.01$ & $12.28\pm 0.04$ & $15.12\pm  0.01$ & $13.64\pm 0.04$ \\
$\rm PS1/\textit{z}$ & $8658$ & $14.07\pm 0.01$ & $12.04\pm 0.02$ & $12.04\pm 0.05$ & $14.95\pm  0.02$ & $13.46\pm 0.04$ \\
$\rm PS1/\textit{y}$ & $9603$ & $13.8\pm 0.03$ & $11.8\pm 0.03$ & $11.83\pm 0.04$ & $14.67\pm  0.05$ & $13.23\pm 0.04$ \\
2MASS/J & $16620$ & $13.69\pm 0.04$ & $11.56\pm 0.03$ & $11.57\pm 0.02$ & $14.68\pm  0.05$ & $12.87\pm 0.03$ \\
2MASS/H & $12482$ & $13.37\pm 0.03$ & $11.36\pm 0.03$ & $11.34\pm 0.02$ & $14.43\pm  0.05$ & $12.71\pm 0.04$ \\
2MASS/K & $21590$ & $13.55\pm 0.04$ & $11.55\pm 0.03$ & $11.51\pm 0.02$ & $14.72\pm  0.05$ & $12.83\pm 0.04$ \\
WISE/W1 & $33526$ & $14.32\pm 0.01$ & $12.34\pm 0.01$ & $12.3\pm 0.01$ & $15.25\pm  0.02$ & $13.6\pm 0.02$ \\
WISE/W2 & $46028$ & $14.96\pm 0.02$ & $13.04\pm 0.01$ & $12.94\pm 0.02$ & $15.85\pm  0.02$ & $14.29\pm 0.02$ \\
WISE/W3 & $120000$ & $14.76\pm 0.06$ & $13.56\pm 0.07$ & $13.03\pm 0.03$ & $15.53\pm  0.07$ & $15.49\pm 0.16$ \\
WISE/W4 & $220000$ & $16.19\pm 0.82$ & $13.82\pm 0.23$ & $12.39\pm 0.04$ & $16.15\pm  0.56$ & $>14.56$\\
\enddata
\centering
\tablenotetext{a}{All measurements are reported in the AB system and are corrected for Galactic line of sight reddening.}

\end{deluxetable*}


%

\begin{deluxetable*}{ccccccc}[h]
\centering
\tablecaption{Photometry of SN host-galaxies analyzed in this work (continued)}
\tablewidth{26pt} 
\tablehead{\colhead{Instrument/Filter} & \colhead{$\lambda_{eff}$ ($\AA$)} & \colhead{PTF16pq} & \colhead{SN\,2016hil$^{b}$} & \colhead{SN\,2018fsh} & \colhead{SN\,2019ape} & \colhead{SN\,2020uik}}
\tabletypesize{\scriptsize} 
\startdata 
GALEX/FUV & $1542$ & $19.04\pm  0.11$ & $19.83\pm  0.08$ & $19.47\pm  1.43$ & $17.67\pm  0.05$ & $19.77\pm  0.15$ \\
GALEX/NUV & $2274$ & $19.36\pm  0.07$ & $20.42\pm  0.17$ & $18.35\pm  0.47$ & $16.57\pm  0.02$ & $18.36\pm  0.06$ \\
SDSS/\textit{u} & $3595$ & $16.25\pm  0.09$ & $17.67\pm  0.08$ & $16.67\pm  0.12$ & $14.89\pm  0.03$ & - \\
SDSS/\textit{g} & $4640$ & $14.49\pm  0.02$ & $15.88\pm  0.03$ & $15.13\pm  0.04$ & $13.36\pm  0.01$ & - \\
SDSS/\textit{r} & $6122$ & $13.66\pm  0.01$ & $15.02\pm  0.02$ & $14.34\pm  0.04$ & $12.59\pm  0.01$ & - \\
SDSS/\textit{i} & $7440$ & $13.22\pm  0.02$ & $14.57\pm  0.02$ & $13.93\pm  0.03$ & $12.2\pm  0.01$ & - \\
SDSS/\textit{z} & $8897$ & $13.0\pm  0.03$ & $14.2\pm  0.04$ & $13.62\pm  0.04$ & $11.91\pm  0.02$ & - \\
$\rm PS1/\textit{g}$& $4776$ & $14.43\pm  0.02$ & $15.84\pm  0.03$ & $15.08\pm  0.02$ & $13.47\pm  0.01$ & $15.71\pm  0.03$ \\
$\rm PS1/\textit{r}$ & $6130$ & $13.7\pm  0.01$ & $15.09\pm  0.02$ & $13.97\pm  0.02$ & $12.74\pm  0.02$ & $15.09\pm  0.03$ \\
$\rm PS1/\textit{i}$ & $7485$ & $13.31\pm  0.01$ & $14.64\pm  0.01$ & $14.34\pm  0.02$ & $12.46\pm  0.01$ & $14.76\pm  0.03$ \\
$\rm PS1/\textit{z}$ & $8658$ & $13.09\pm  0.01$ & $14.47\pm  0.01$ & $13.79\pm  0.04$ & $12.13\pm  0.01$ & $14.57\pm  0.07$ \\
$\rm PS1/\textit{y}$ & $9603$ & $12.85\pm  0.02$ & $14.16\pm  0.03$ & $13.57\pm  0.07$ & $11.86\pm  0.02$ & $14.3\pm  0.08$ \\
2MASS/J & $16620$ & $12.59\pm  0.05$ & $14.14\pm  0.05$ & $13.32\pm  0.04$ & $11.63\pm  0.02$ & $14.07\pm  0.04$ \\
2MASS/H & $12482$ & $12.27\pm  0.05$ & $13.85\pm  0.05$ & $13.18\pm  0.04$ & $11.49\pm  0.03$ & $14.05\pm  0.05$ \\
2MASS/K & $21590$ & $12.56\pm  0.05$ & $14.03\pm  0.05$ & $13.37\pm  0.05$ & $11.65\pm  0.02$ & $14.28\pm  0.07$ \\
WISE/W1 & $33526$ & - & $14.49\pm  0.01$ & $14.08\pm  0.02$ & $12.36\pm  0.01$ & $15.0\pm  0.02$ \\
WISE/W2 & $46028$ & $14.01\pm  0.02$ & $15.13\pm  0.03$ & $14.78\pm  0.02$ & $13.0\pm  0.01$ & $15.69\pm  0.02$ \\
WISE/W3 & $120000$ & $14.54\pm  0.10$ & $16.2\pm  0.12$ & $16.12\pm  0.45$ & $12.98\pm  0.06$ & $17.16\pm  0.26$ \\
WISE/W4 & $220000$ & $14.69\pm  0.57$ & $14.9\pm  0.12$ & $>14.83$ & $13.2\pm  0.19$ & $16.35\pm  0.73$ \\
\enddata
\centering
\tablenotetext{a}{All measurements are reported in the AB system and are corrected for Galactic line of sight reddening.}
\tablenotetext{b}{W3 photometry is estimated from the W2-W3 ALLWISE color of \cite{Irani2019b} (see text).} 
\vspace{-0.15cm}
\end{deluxetable*}
%
%

\section{Results}
\label{sec:results}
In this section we analyze the properties of the three newly-discovered SNe. In Sec. \ref{subsec:18fsh20uik} we discuss the light curve, spectra and location of the SNe II \fsh\ and \uik. In Sec. \ref{subsec:19ape} we analyze our observations of \ape. We argue it is a typical SN Ic by comparing it to other SNe Ic, and by modelling its spectroscopic evolution. We present evidence that the SN has formed near the explosion site and is a result of low-level star formation in its elliptical host. In Sec.  \ref{subsec:sample_properties} we derive host galaxy properties for the combined sample of CCSNe in ellipticals and compare them to the hosts of all BTS SNe (CCSNe and SNe Ia). We discuss these results and their implications for star formation in elliptical galaxies in \ref{sec:discussion}. 

\subsection{\fsh\ and \uik}
\label{subsec:18fsh20uik}
\subsubsection{Light curve properties}
 \uik\ is a SN IIP detected early in its plateau phase at an absolute magnitude of $M_{\textit{r}} = -16.9\ \rm mag$, which we adopt as the peak luminosity. The plateau continues for 65 days until the SN begins to fade. \fsh\ is a SN IIL - it was detected by ATLAS near peak at a luminosity of $\textit{c} = -18.1\ \rm mag$ and was observed again during its linear decline, $77\ \rm days$ after the estimated time of maximum light. It continues to decline for an additional $150\ \rm days$ until becoming undetectable.

\begin{figure*}
\centering

\includegraphics[width=1.8\columnwidth]{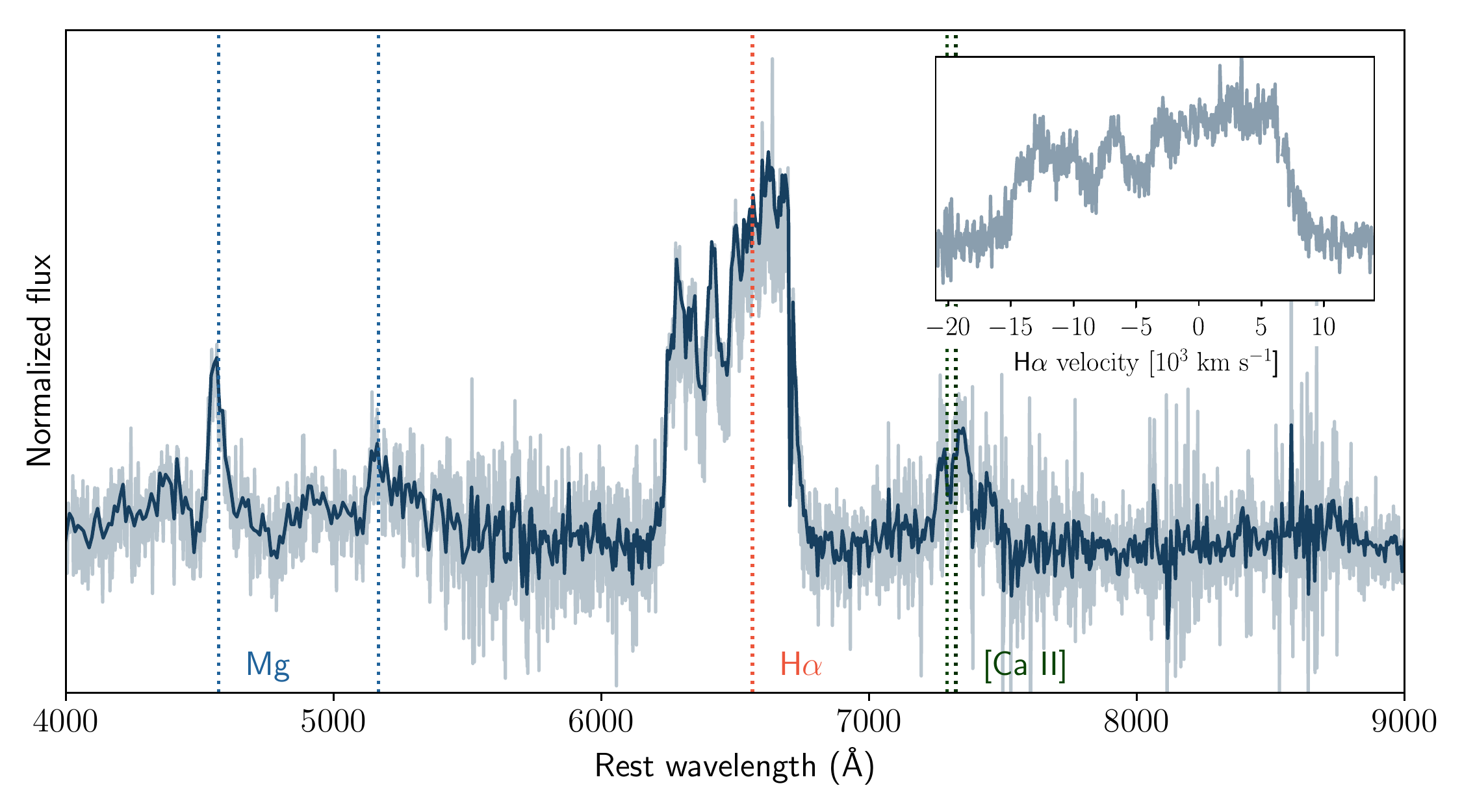}
\caption{The nebular spectrum of SN\,2018fsh. Dashed colored lines denote major features. Inset panel shows a zoom in of the H$\alpha$ feature with wavelength in units of velocity relative to the rest frame. The dark curve is a binned version of the spectrum to guide the eye.} 
\label{fig:18fsh_nebular}

\end{figure*}

\subsubsection{Spectra}
The spectral observations of \fsh\ and \uik\ show that both appear to be spectroscopically normal SNe II. \fsh\ was observed late in its evolution, and shows a typical spectrum for a SN IIL at this phase. We show a comparison with the closest \package{Superfit} match SN\,2014G \citep{deJaeger2019} in Fig. \ref{fig:18fsh_spec}. The spectrum shows strong \Ha\ and moderate \CaII\ emission. We interpret the absorption feature at ~$\lambda5900$ as Na $\lambda5890$ and not He $\lambda5876$ in the absence of additional features at $\lambda6678$ and $\lambda7065$ \citep{galyam2017}. Figure \ref{fig:18fsh_nebular} shows a later nebular spectrum with Mg and \CaII\ emission lines as well as a complex \Ha\ emission profile. The \Ha\ feature has a broad and boxy profile, with velocities ranging from $-15,000$ to 10000 \kms. Similar profiles have been previously attributed to interaction of the fast outer ejecta with CSM \citep{Filippenko1994,Patat1995}. This interpretation is supported by the flattening of the light curve at late times, e.g., as observed in all bands for SN\,1993J \citep{Jerkstrand2015}.

\uik\ has a more detailed spectroscopic sequence with a typical \Ha\ P-Cygni profile, Na I absorption at $\lambda 5890$ as well as Fe II absorption lines ($\lambda4924$, $\lambda5018$\ and $\lambda5169$) and with typical \CaII\ emission developing over time. Neither of the two events shows significantly strong Ca emission compared to other SNe II to be considered peculiar.

\subsubsection{Limits on an underlying host galaxy}
\begin{figure*}
\centering
\includegraphics[width=1.6\columnwidth]{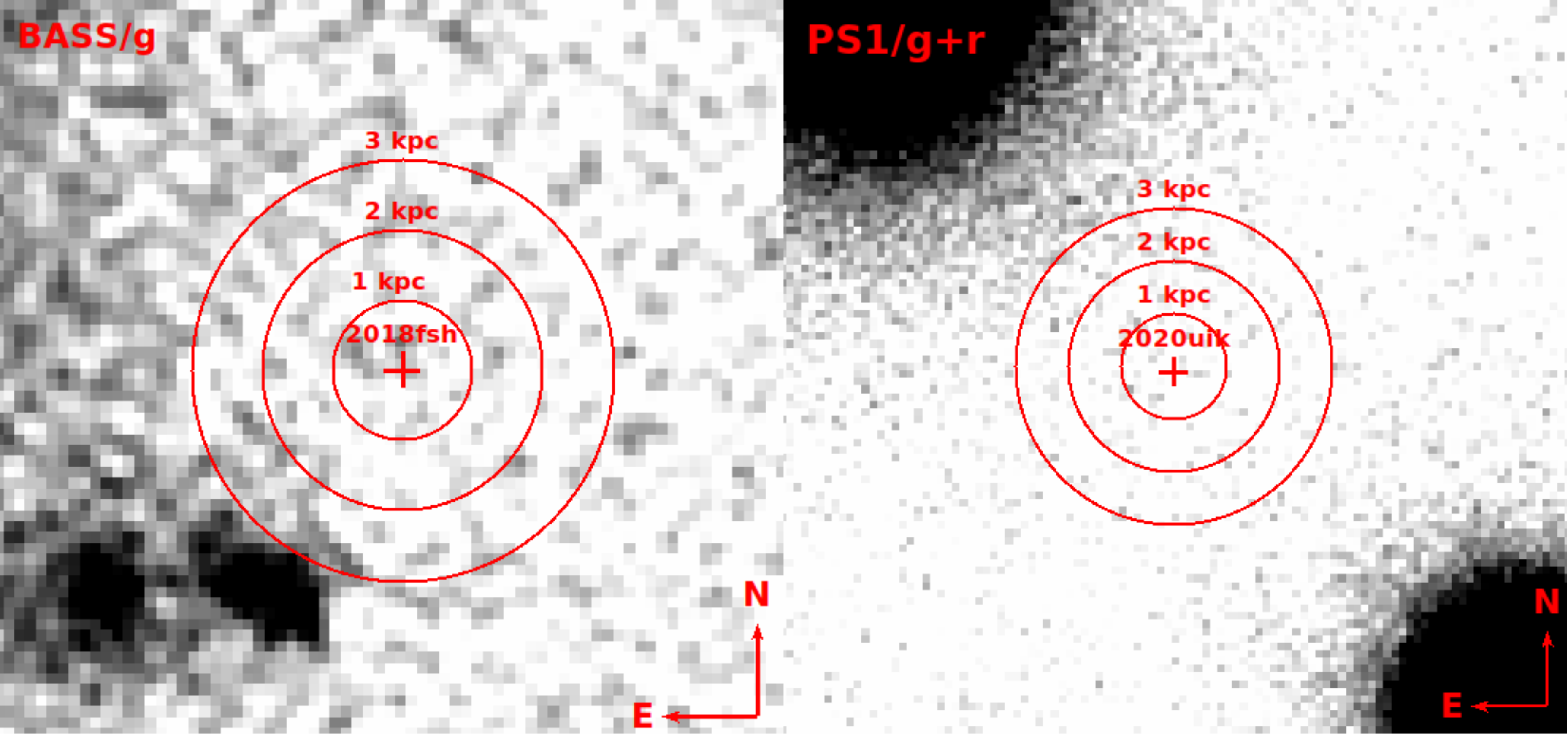}

\caption{Deep archival images of the explosion sites of \fsh\ and \uik. PS1 \textit{g} and \textit{r} images were stacked to increase the signal at the site of \uik. Concentric circles at 1, 2, 3 kpc are shown around the SN position.}  
\label{fig:deep_host}

\end{figure*}

We derive limits on the presence of a compact source at the sites of SN 2018fsh and SN 2020uik using deep archival imaging as described in \cite{Irani2019b}. For \fsh\ we use Legacy Survey images \citep{Dey2019} from the Beijing-Arizona Sky Survey fields \citep{Zou2017}, and for \uik\ we use deep PS1 imaging \citep{Flewelling2020}. Our $5\sigma$ limits are $m_{\textit{g}}>24.69\ \rm mag$ at the site of \fsh\ and  $m_{\textit{g}}>23.45\ \rm mag$ at the site of \uik\, corresponding to luminosity limits of $M_{\textit{g}}=-10.73\ \rm mag$ and $M_{\textit{g}}=-11.91\ \rm mag$ respectively. In Fig. \ref{fig:deep_host} we show cutouts of the explosion sites. In both cases, there are no sources within a few kpc of the SN location. 

\subsection{\ape }
\label{subsec:19ape}

\subsubsection{Light-curve properties}
The light curves of \ape\ are shown in Fig. \ref{fig:19ape_lc}. \ape\ rose to an \textit{r}-band peak luminosity of $M_{\rm peak}=-16.75\pm 0.05\ \rm mag$, $17.1\pm0.7$ days after the explosion in the SN rest frame. The time of peak is determined by fitting a 3rd-order polynomial to the \textit{r}-band data around the peak (10-30 days after $t_{exp}$), and the error is estimated by varying the fit range and order. The decline of the light curve is observed over the course of $~100$ days, with a re-brightening of the \textit{g}- and B-band light curves observed starting from $t = 60$ days and reaching an unusual secondary peak of $\textit{g} = -14.6\ \rm mag$  at $t \sim 80$ days. The transient is then observed again at $t = 305$ days at a luminosity of $\textit{r} = -13.0\pm{0.2}\ \rm mag$ before fading below the detection limit. The late-time decline of $2.5\ \rm mag$ over the course of 250 days is consistent with  \Cofs\  decay (0.98 mag per 100 days; \citealt{Woosley1989}), but the exact \Nifs\ mass cannot be measured directly due to the poor light-curve sampling at late times.

\subsubsection{Spectroscopic comparison of SN 2019ape and other SNe Ic}
\label{subsec:2019ape_comp}

\begin{figure*}
\centering
\includegraphics[width=2\columnwidth]{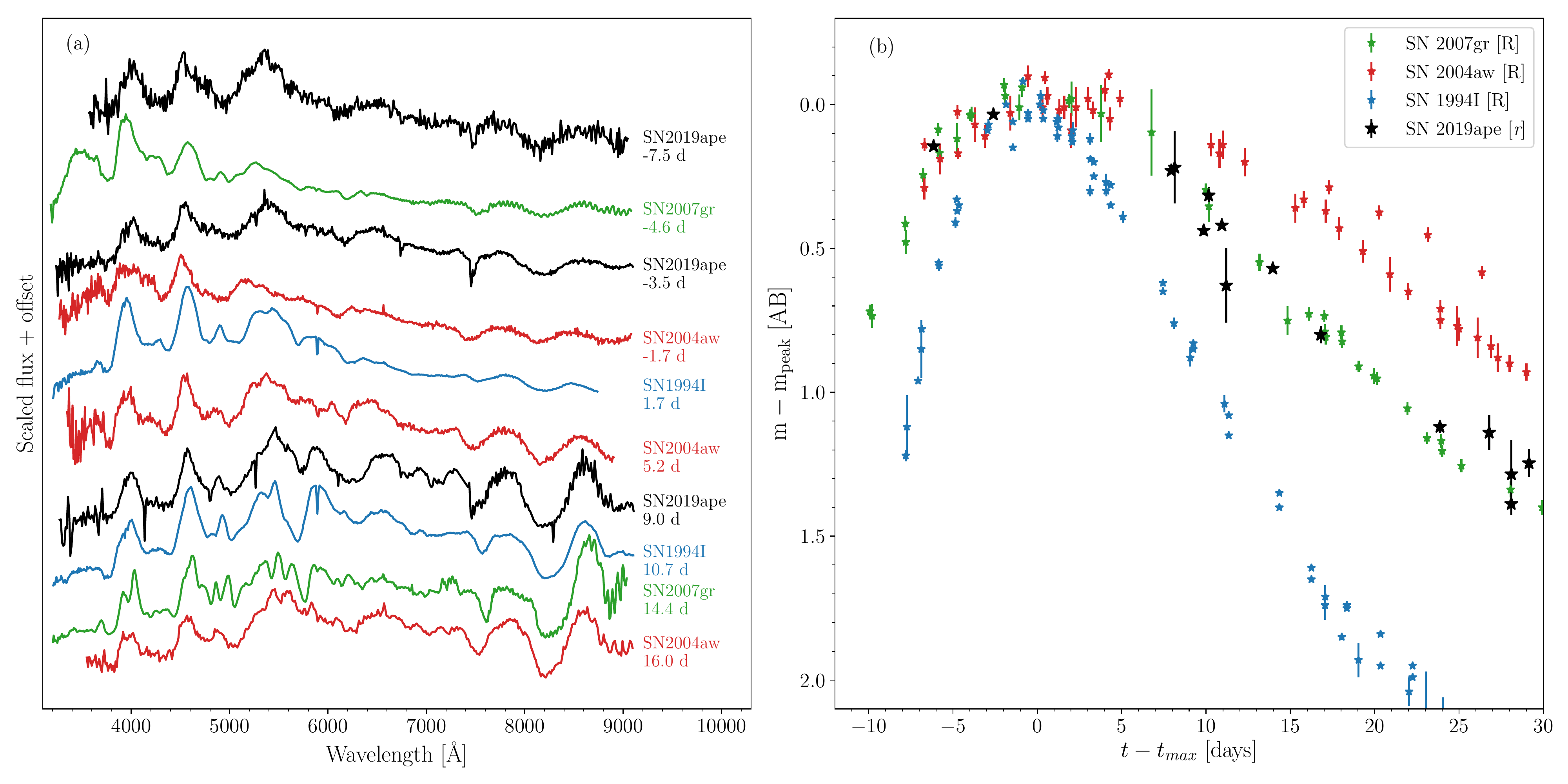}
\caption{Comparison of the (a) spectra  and (b) \textit{r}/R-band light-curves of SN\,2019ape to those of SN\,2007gr (green), SN\,2004aw (red) and SN\,1994I (blue). The spectra are corrected for both host and Milky-Way extinction as described in the text.} 
\label{fig:19ape_comp}

\end{figure*}

Qualitatively, the spectra of SN\,2019ape are similar to those of SNe Ic.
In Fig. \ref{fig:19ape_comp} the spectra are compared with the well-observed SNe Ic SN\,1994I \citep{Filippenko1995}, SN\,2004aw, the best fitting SN to \ape\ with \package{Superfit} \citep{Taubenberger2006}, and SN\,2007gr \citep{Valenti2008,Modjaz2014} around maximum light and at two weeks past maximum.  
Around maximum light, the spectroscopic features are similar in position and in strength among the objects, in particular those associated with \FeII\ and \MgII\ in the region bluewards of 5500 \AA.  
Of note is that the spectrum of SN\,2019ape at $-7.5$ days peaked at around 5500 \AA. This is unlike the early spectra of SNe 2004aw and 2007gr which are bluer despite being observed later in their evolution (e.g., the spectrum of SN\,2007gr at $-4.6$ days peaks at $\sim4000$ \AA).
Over the the course of a week, the spectra of the comparison objects gradually redden as they reach maximum light and the ejecta cool.
The SN\,2019ape spectrum at $-3.5$ d is similar in this respect to the comparison objects around peak or slightly after. The spectrum is a very good match to the spectrum of SN\,2004aw at around $+5$ days.
As time progresses the objects gradually evolve to look similar, although SN\,2019ape does not display the narrow lines of SN\,2007gr and retains the slight blending of the \FeII\ lines in the blue part of the spectrum as seen in SNe 1994I and 2004aw \citep[For a discussion of this, see][]{Prentice2017}.
It can be concluded that SN\,2019ape is a normal SN Ic supernova, albeit with some differences in its pre-peak spectra, and the peculiar secondary peak in \textit{g}-band.

\subsubsection{Photospheric-phase spectral modelling}
\label{subsec:SN2019ape_modelling}
We model the spectra of SN\,2019ape with a 1D radiative transfer code which has been used extensively to model the spectra of Stripped-Envelope SNe (SE-SNe) \citep[e.g.,][]{Mazzali2000b,Mazzali2002,Mazzali2006, Sauer2006,Prentice2018,Ashall2019,Teffs2021}.
The code is described in detail by \cite{Mazzali1993,Lucy1999,Mazzali2000}, and is the source of a parameter study by \cite{Ashall2020}. The model requires an input density profile, ejecta composition, photospheric velocity, epoch, and luminosity which it then uses to approximate the expanding SN ejecta, which are assumed to be in homologous expansion. 
Radiation is assumed to be emitted at a sharp inner boundary (the ``the inner photosphere''), following a blackbody distribution with temperature $T_\mathrm{bb}$, which is found through an iterative MC process.
The code follows the propagation of `photon packets' through the SN atmosphere, as represented by the density and abundance profiles. These packets can be subject to Thomson scattering and line absorption in the model ejecta, both fluorescence and reverse fluorescence are possible. The interaction of photons and the gas redistributes the temperature and determines the ionization and excitation states of the gas, which are computed self-consistently using the nebular approximation \citep{AbbottLucy1985,Pauldrach1996}. Finally, the emergent spectrum is found by calculating the formal integral of the radiation field.

\begin{figure*}
\centering
\includegraphics[width=2\columnwidth]{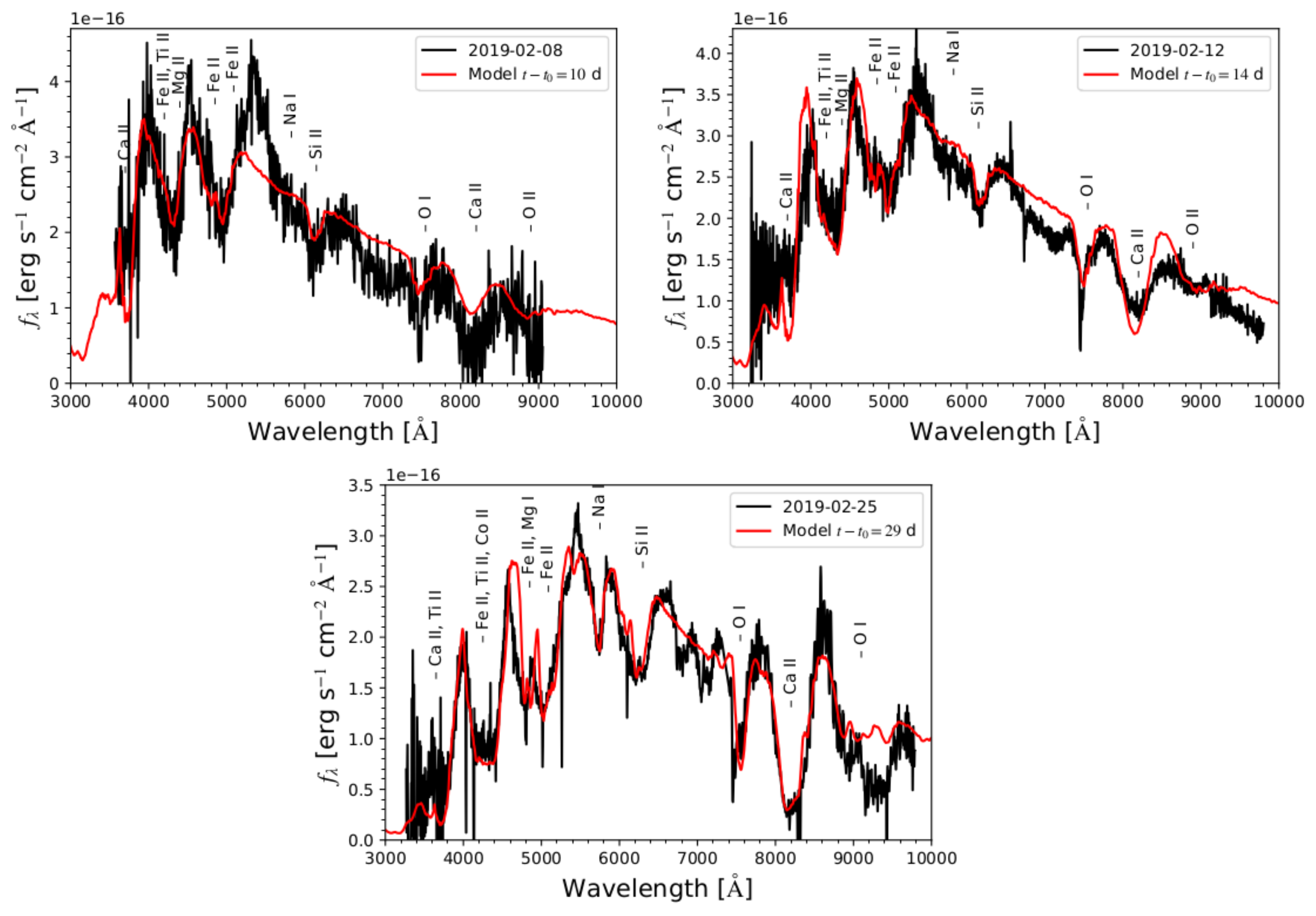}
\caption{Best fit model spectra for the selected spectra of \ape.}
\label{fig:best_fit}
\end{figure*}

We modelled three spectra of \ape, one prior to \textit{r}-band maximum, one at maximum, and one a few weeks later. As \ape\ shows spectral similarities to SN\,2004aw, we started with the density profile and abundance distribution from the model found in \citet{Mazzali2017}, which in turn is based on the CO21 1D hydrodynamical model developed for SN\,1994I \citep{Nomoto1994}. This initial model was modified using the scaling relations from \citet{Hachinger2009} for the ejected mass \mej\ and the explosion kinetic energy \ek\ until a model that reasonably reproduces the spectral flux and line profiles is found, and then the abundances are iteratively modified until the bulk of the features are well reproduced.
As we did not model the light curve in parallel with the spectra, the values of \mej\ and \ek\ that we determined are approximate. Additionally, the lack of early-time spectra and photometry limit the accuracy of the determination of \ek\ using this method (see \citealt{Mazzali2017}).
Our modelling suggests that a good fit is obtained using a density profile with \mej\ $=2$ \msun\ and \ek\ $=2\times 10^{51}\ \rm erg$, and a specific kinetic energy that is similar to SN\,2004aw ($E_{k,51}$/\mej\ $ \approx 1$).
As the first modelled epoch is approximately 10 days after \texp\ as estimated in Sec. \ref{sec:det_ape}, we used two dummy shells to model the high velocity material properly. As we have no early spectra, these two shells were primarily used to reduce high-velocity line formation of certain elements. In addition, we use an additional dummy shell between the second and third modelled epochs ($t \sim 14$ days and $t \sim 27$ days respectively) as this is an extensive gap in spectral evolution. The resulting models are shown in Fig. \ref{fig:best_fit} with the abundances of each shell given in Table \ref{tab:modelling}.

The first modelled spectrum was obtained on 2019 Feb. 8, $\sim 10$ days from estimated explosion time and $\sim 7.6$ days prior to r-band peak. For this spectrum we find \vph\ $=12500$ \kms\ and $L = 1.40\times10^{42}$ \ergs. 
The model manages to capture many of the spectral features well, with a few issues (see first panel of Fig.~\ref{fig:best_fit}).
The abundance at this phase is primarily carbon ($\sim$66\%), with $\sim30$\% neon and only 3\% oxygen. A larger oxygen abundance leads to an excessively strong \OI\ $\lambda 7774$ feature in the synthetic spectrum. In the dummy shells above this photosphere, the oxygen mass is reduced even further, leaving a C and Ne rich outer shell.
Comparatively, at this epoch the models shown in \cite{Mazzali2017} had approximately equal abundances of carbon and oxygen (35\%).
The observed \OI\ $\lambda 7774$ line in this epoch is also noisy and likely contaminated with a telluric line, which makes it difficult to fit.
As mentioned previously, a key difference between SN\,2004aw and \ape\ is that the spectrum of \ape\ is redder and peaks at $\sim5500$ \AA\ rather than $\sim4500$ \AA.
Two noticeable issues with the model are the inability to replicate the peak near 5500\,\AA\ and the weak NIR \CaII\ line. 
The peak could be driven by re--emission from the iron features near 5000\,\AA\ or possibly other Fe--group elements, however the high abundances of Fe or Fe--group elements needed to drive the flux upwards in this region produce absorption features that are far too strong as well as other lines that are not observed. Increasing the total luminosity can also reproduce the peak at 5500\,\AA\ but produces far too high a flux level at wavelengths redder than 6000\,\AA.
The second issue is that the \CaII\ NIR feature is too weak, which is likely due to the model having too high a temperature at this phase, leading to the over-ionization of Ca. Lowering the temperature produces a worse fit to other parts of the spectrum, and increasing the Ca abundance is inconsistent with the later spectra.

\begin{deluxetable*}{lcccccccccccc}
\label{tab:modelling}
\centering
\tablecaption{Parameters and selected abundance fractions for the model shells.}

\tablewidth{20pt} 

\tablehead{\colhead{$t$ [days]} & \colhead{\vph\ [\kms]} & \colhead{$T_{bb}$ [K]} & \colhead{C} & \colhead{O} &\colhead{Ne} &\colhead{Na} & \colhead{Mg}& \colhead{Si}& \colhead{S}& \colhead{Ca}& \colhead{$^{56} \rm Fe$}& \colhead{$^{56} \rm Ni$}} 
\tabletypesize{\scriptsize} 

\startdata 
Dummy shell & 23000 & - & 0.7 & 0.004 & 0.29 & 0.0 & 0.005 & 0.0005 & 0.00025 & 0.0 & 0.0001 & 0.0001 \\
Dummy shell & 17000 & - & 0.7 & 0.0065 & 0.29 & 0.0 & 0.0025 & 0.0005 & 0.00025 & $3\times 10^{-6}$ & 0.0001 & 0.0001 \\
10 & 12500 & 7249.6 & 0.66 & 0.035 & 0.3 & 0.0005 & 0.0025 & 0.0015 & 0.0006 & $7\times 10^{-6}$ & 0.0001 & 0.0005 \\
14 & 10500 & 7028.2 & 0.5 & 0.2 & 0.3 & 0.00055 & 0.001 & 0.0022 & 0.002 & 0.0 & 0.0005 & 0.002 \\
Dummy shell & 8500 & - & 0.5 & 0.3 & 0.15 & 0.001 & 0.001 & 0.015 & 0.035 & $2\times 10^{-6}$ & 0.0025 & 0.0025 \\
27 & 6500 & 6327 & 0.5 & 0.3 & 0.02 & 0.002 & 0.001 & 0.015 & 0.05 & $1\times 10^{-6}$ & 0.0025 & 0.05 \\
\enddata

\end{deluxetable*}

We next model a spectrum that was obtained four days later, on 2019 Feb. 12 ($\sim$14 days after the explosion and 2.8 days before the peak). The input parameters were \vph\ $=11000$\,\kms\ and $L = 1.64\times10^{42}$ \ergs. This spectrum is quite similar to the one at $t-t_{\rm exp} \sim 10$\,d but is marginally redder, suggesting a lower temperature. The abundances are similar to those of the previous spectrum, as the velocities only differ by 1500\,\kms\ and therefore represent shells located close within the ejecta profile.
Although the main features of the spectrum are replicated, owing to the lower temperature the \CaII\ NIR and H\&K features are stronger in this model than in the previous one.

The final spectrum modelled was obtained on 2019 Feb. 25, $\sim 27$ days after the explosion and 9.4 days after the peak. Here, we used \vph\ $=6500$\,\kms\ and $L = 1.23\times10^{42}$ \ergs. 
As before, many of the features in the spectrum are replicated by the model, apart from the peak at 5500\,\AA, which shows an \FeII\ line with a rest wavelength of 5534\,\AA\ that is clearly not visible in the observed spectrum. 
Comparing to the other SNe of similar epochs in Fig. \ref{fig:19ape_comp}, the line is seen in SN\,2004aw, SN\,2007gr, and possibly in SN\,1994I. However, the Fe abundance is required to reproduce other features, suggesting that the Fe abundance at higher velocities may be too low. But, as discussed previously, increasing the Fe abundance in the outermost regions produces unwanted lines in those epochs.
Several lines are observed in the spectrum in the region between 6600 and 7100\,\AA\ that are not reproduced by the model. These are often blends of \CI\ and Fe--group elements. Carbon is still mostly singly ionized, and the issues with the Fe distribution still holds.
The \OI\ abundance is still low during this phase, but the \OI\,7774\,\AA\ line is saturated, and is less responsive to abundances changes at this epoch. The absorption near 9300\,\AA\ is possibly due to strong \CI\ features that are only partially replicated.\\

\subsubsection{Star-formation of the host galaxy}
\label{subsec:SN2019apeSF}
\begin{figure*}
\centering
\includegraphics[width=2\columnwidth]{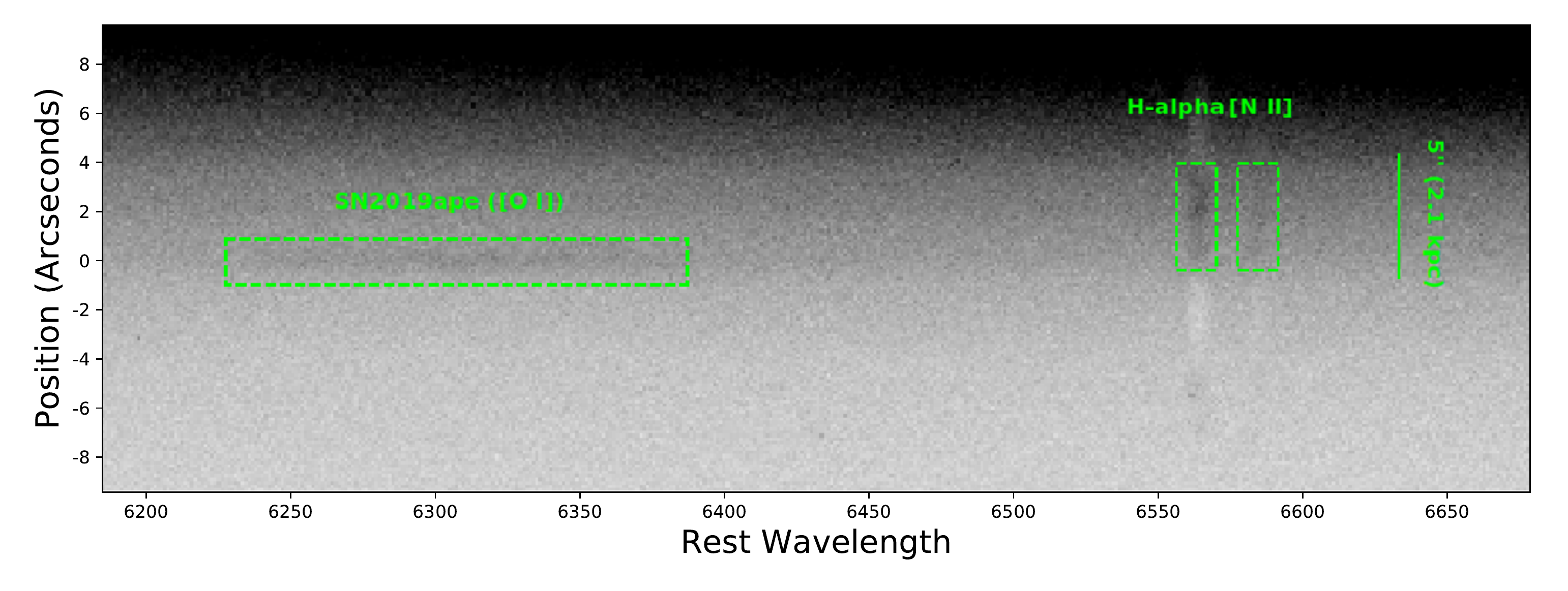}
\caption{2D cutout of the nebular spectrum of SN\,2019ape. The trace of the SN is faint and becomes significant relative to the galaxy background only at the [O I] nebular feature. Both the \Ha\ and weaker [\ion{N}{2}] narrow lines extend to the SN site, indicating an underlying star forming region. The 2D frames are sky subtracted using the sky-subtraction routine of \package{LPipe}, which removes some of the host-galaxy emission lines for larger galaxies.}  
\label{fig:local_sf}
\end{figure*}

Massive stars are associated with regions of elevated star formation which can be traced by UV continuum and \Ha\ emission \citep{Kennicutt1998a}. We here consider both star formation in the local environment of \ape\ and the total star formation rate of its host galaxy. \cite{De2020} acquired a deep nebular spectrum of \ape\  $\sim300$ days after the explosion. The 2D spectrum reveals a compact H$\alpha$ emitting region in close proximity to the SN site on top of the narrow absorption feature observed throughout the slit. In Fig. \ref{fig:local_sf}, we show a cutout of the 2D image with the trace of \ape\ and the H$\alpha$ emitting region highlighted. Using \package{SAOImageDS9} \citep{Joye2003} we measured the strength of the \Ha\ emission relative to the local absorption background. After correcting for MW extinction we find a flux of $f_{\rm H\alpha} =1.4\pm0.5\times 10^{-16}\ \rm erg\ s^{-1}\ cm^{-2}$ and an integrated \Ha\ luminosity of $3.1 \pm 1\times 10^{38}\ \rm erg\ s^{-1}$ - a typical value for an H II region \citep{Kennicutt1989}. We convert this to a star-formation rate (SFR) using the calibration by \cite{Kennicutt1998a} and find that this regions forms $2.4\times 10^{-3} M_{\odot}\ \rm yr^{-1}$.   Globally the host-galaxy is relatively UV bright, with a color of  $NUV - \textit{r}_{\rm PS1}=4.03\pm0.03\ \rm mag$ indicating it has undergone an episode of recent ($< 100$ Myr) star-formation.  (\citealt{suh2011}, \citealt{Yi2005}).

\subsection{Host-galaxy sample properties}
\label{subsec:sample_properties}

\begin{deluxetable*}{llcc}
\label{tab:host_params}
\centering
\tablecaption{SN host-galaxy properties derived in this work}

\tablewidth{20pt} 

\tablehead{\colhead{SN} & \colhead{Host} & \colhead{$SFR\ (M_{\odot}\ \rm yr^{-1})$}  & \colhead{Mass ($10^9\ M_{\odot}$) }} 
\tabletypesize{\scriptsize} 

\startdata
SN2003ky & NGC 4001 & $1.18 \pm 0.09$ & $37.5 \pm 1.0$ \\
SN2006ee & NGC 774 & $0.17 \pm 0.01$ & $19.6 \pm 0.5$ \\
SN2006gy & NGC 1260 & $0.18 \pm 0.04$ & $36.4 \pm 1.0$ \\
PTF10gqf & WISEA J150350.31+553738. & $0.26 \pm 0.01$ & $13.9 \pm 0.4$ \\
PS1-12sk & RXC J0844.9+4258 & $1.62 \pm 0.09$ & $98.7 \pm 3.0$ \\
SN2016hil & WISEA J011024.51+141238. & $0.26 \pm 0.05$ & $50.9 \pm 1.9$ \\
PTF16pq & CGCG 091-056 & $0.16 \pm 0.01$ & $32.7 \pm 0.9$ \\
SN2018fsh & MCG +07-18-013 & $0.43 \pm 0.19$ & $15.2 \pm 0.4$ \\
SN2019ape & NGC 3426 & $1.22 \pm 0.02$ & $49.0 \pm 1.3$ \\
SN2020uik & WISEA J080154.84-064527. & $0.4 \pm 0.02$ & $5.6 \pm 0.2$ \\
- & Galaxy Zoo ellipticals$^{\dagger}$ & $0.09^{+0.09}_{-0.05}$ & $8.27^{+19.33}_{-5.79}$ \\
- & Galaxy Zoo spirals$^{\dagger}$     & $0.43^{+0.78}_{-0.28}$ & $2.43^{+5.69}_{-1.71}$ \\
\enddata
\tablenotetext{\dagger}{\ Values quoted here represent the sample mean and standard deviation.}
\end{deluxetable*}

\begin{figure*}
\centering
\includegraphics[width=2\columnwidth]{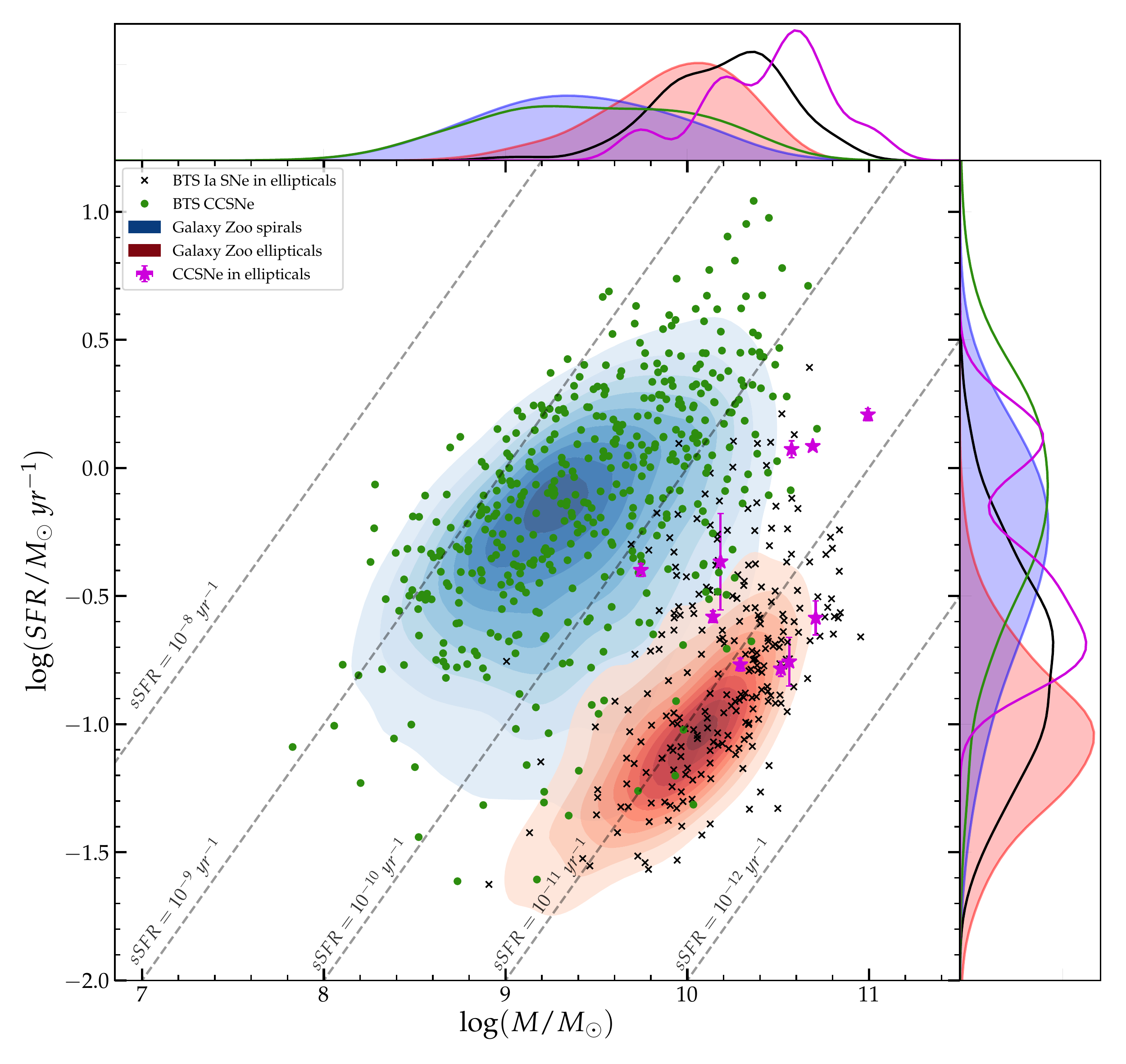}
\caption{SFR and stellar masses for (1) BTS SNe Ia in elliptical galaxies (black crosses), (2) BTS CCSNe (green circles), (3) our sample of CCSNe in ellipticals (magenta stars). Diagonal gray dashed lines are lines of equal sSFR, and red and blue contours correspond to Galaxy Zoo ellipticals (which satisfy the color criteria described in Sec. \ref{sec:sample}) and Galaxy Zoo spirals. The top and right panels show the corresponding 1D kernel density estimates using the same color scheme as the 2D plot.}
\label{fig:sfr_mass} 
\end{figure*}

We use the host photometry of the objects in our sample and the archival photometry for the BTS SN hosts to derive their SFR and stellar masses. Traditionally, SFR can be derived using UV luminosities that are dominated by massive stars (e.g. \citealt{Kennicutt1998a}, \citealt{Salim2007}). However, this tracer is sensitive to dust attenuation (e.g. see \citealt{Calzetti1995}, \citealt{Buat1999}). To compensate for the effects of dust attenuation, we use the UV SFR indicator calibrated by \cite{Salim2007}. Stellar mass values are estimated from the $W2$ brightness using the calibration of \cite{Wen2013}. We report the derived SFR and stellar mass estimates in Table \ref{tab:host_params}, along with the mean SFR and stellar mass estimates for Galaxy Zoo spirals and ellipticals. In Fig. \ref{fig:sfr_mass} we plot the SFR of BTS CCSNe hosts, our sample of CCSNe in elliptical galaxies and BTS SNe Ia in elliptical galaxies. The elliptical host galaxies of CCSNe show more star formation on average compared to the general elliptical galaxy population -  $0.41^{+0.53}_{-0.23}\ M_{\odot}\ \rm yr^{-1}$ compared to $0.09^{+0.09}_{-0.05}\ M_{\odot}\ \rm yr^{-1}$ for Galaxy Zoo ellipticals and $0.76^{+1.47}_{-0.72}\ M_{\odot}\ \rm yr^{-1}$ for the general CCSN host-galaxy population. CCSNe in elliptical galaxies are also more massive than most elliptical galaxies, with $2.76^{+3.24}_{-1.49} \times 10^{10} M_{\odot}$ compared to an average of $0.83^{+1.95}_{-0.58} \times 10^{10} M_{\odot}$ for the Galaxy Zoo ellipticals. The high SFR of CCSNe elliptical hosts can be explained by their higher mass. The specific SFR (sSFR; i.e. the SFR per unit mass) of CCSNe ellitpicals hosts is $1.49^{+2.08}_{-0.87}\times 10^{-11}\ \rm yr^{-1}$ compared to $1.06^{+1.03}_{-0.52}\times 10^{-11}\ \rm yr^{-1}$ for galaxy zoo ellitpicals, and $2.60^{+4.99}_{-2.54}\times 10^{-10}\ \rm yr^{-1}$ for the hosts of BTS CCSNe.

\section{Discussion}
\label{sec:discussion}
In this paper, we have presented and analyzed three CCSNe occurring in elliptical galaxies from the ZTF BTS experiment and their hosts. In this section we will discuss the implications of our results regarding star formations in ellipticals, as inferred from the population of CCSNe which they host.

\subsection{CCSNe as representative members of their spectroscopic classes}
Finding a CCSN in an elliptical galaxy can be either a sign of residual star formation, or that the progenitor of the event in question was not a massive star. The most recent example of the later are the class of Ca-rich Type Ib supernovae\citep{Perets2010}. While these transients have spectra consistent with SNe Ib near maximum light, they occur predominantly in passive environments and galaxy outskirts, which argues against a massive star origin (e.g. \citealt{Perets2010}, \citealt{Kasliwal2012},  \citealt{Lunnan2017}, and most recently \citealt{De2020}). In addition to their remarkable locations, Ca-rich SNe display peculiar features - strong Ca emission lines in their nebular phase, and a lower luminosity compared to typical SNe Ib. 

A non-massive-star origin has been suggested for individual CCSNe offset from elliptical galaxies, such as (1) for PS1-12sk \citep{sanders2013,Hosseinzadeh2019}, where deep \textit{HST} UV imaging excludes local star formation, and (2) for SN\,2016hil \citep{Irani2019b}, where a double peaked light-curve and a low metallicity spectrum were observed. However, conclusions regarding the (potentially) peculiar properties of CCSNe in elliptical host galaxies were difficult to reach based on isolated events, and due to noisy and sparsely sampled photometry and spectroscopy in the case of SN\,2016hil. 

\begin{figure*}
\centering
\includegraphics[width=1.4\columnwidth]{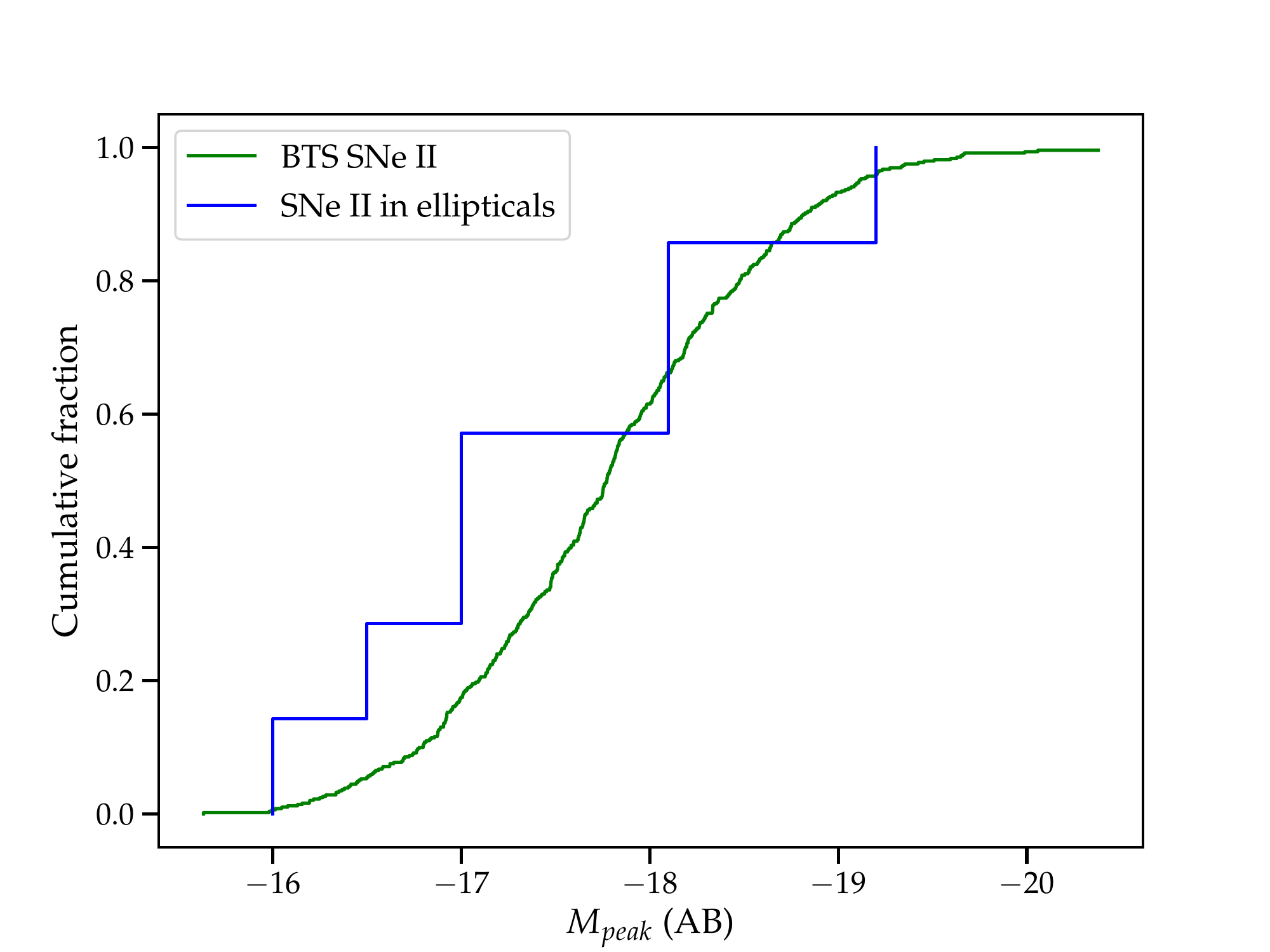}
\caption{The peak \textit{r}-band absolute magnitude cumulative distributions for SNe II in elliptical galaxies (blue curve) compared to SNe II from the BTS survey (green curve).}
\label{fig:peak_mag} 
\end{figure*}

Here, we consider the combined properties of SNe II in ellipticals, which we can now study as a population. Figure \ref{fig:peak_mag} shows the peak luminosities of Type II SNe analyzed in this work, compared to the peak-luminosity distribution of BTS SNe II in our comparison sample. With only 7 SNe II, our sample is too small for a meaningful two-sample Kolmogorov-Smirnov (KS) test of the two peak luminosity distributions. We compare the mean and the standard deviations of the two distributions instead. Our sample of SNe II in ellitpicals has a mean peak absolute magnitude of $-17.3 \pm 1.0$ mag, consistent with $-17.8 \pm 0.8$ mag for all spectroscopically regular SNe II in the BTS sample.\\ 

\indent Unfortunately, we do not have enough data to compare the spectral properties of SNe II in ellipticals to the general SN II population. However, we point out that absence of strong Ca emission in the nebular spectrum of \fsh\ and \ape, (compared to Ca-rich SNe Ib/c or Ia; \citealt{De2020}), and the typical spectral evolution of \uik, suggest these two SNe are typical SNe II. The complex \Ha\ profile seen in the nebular spectrum of \fsh\ has a broad and boxy profile, extending to high-velocities. These were previously seen for the well observed SN\,1993J and SN\,1998S and interpreted as signatures of late-time interaction with circumstellar material  \citep{Filippenko1994,Patat1995,Pozzo2004}. \cite{Sollerman2021} discuss SNe with similar nebular \Ha\ extensively. The presence of extended circumstellar material (CSM) around the progenitor of \fsh\ could indicate some SNe II form through exotic formation channels, such as mergers of intermediate-mass stars as outlined in \cite{zapartas2017}, or a common-envelope phase as suggested by \cite{soker2019}. \cite{zapartas2017} estimates that such binary interaction could account for up to 15\% of SNe II. It remains to be seen if such signatures are common for SNe II with elliptical host-galaxies.\\
\indent While our sample only contains a single SN Ic, we demonstrated in Sec. \ref{subsec:2019ape_comp}  that \ape\ is not a unique event in most respects - both by comparison to other events and by our modelling of \ape\ in \ref{subsec:SN2019ape_modelling}. The later indicates that the abundances, ejected mass and kinetic energy are typical of a normal SN Ic when compared to a sample of events \citep{Prentice2016,Prentice2018b}. Its peak luminosity of $M_{\textit{r}}= -16.75\ \rm mag$ is close to the mean peak luminosity in the BTS sample ($-17.3\pm0.5\ \rm mag$) and the light curve evolution is typical \citep{Prentice2018b}. The only unusual aspect is the secondary \textit{g}-band peak. It might indicate interaction with extended CSM at later times, as for example observed by \cite{Benami2014} or \cite{gutierrez2021}. Similarly to \fsh, the presence of extended CSM might indicate binary evolution is responsible for the formation of the progenitor star of \ape. However, given the rest of its properties and the location of the event, we consider it unlikely that \ape\ emerged from an unusual SN Ic progenitor. In particular, the estimated ejected mass, chemical abundance (dominated by unburned carbon) and energy point toward a massive progenitor. We conclude that our sample of events in elliptical galaxies is overall consistent with having properties similar to the general population of CCSNe. 

\subsection{Localized star-formation in elliptical galaxies}
Based on our observations of the SN sites, we can separate the sample of CCSNe in ellipticals into two groups - the SNe which are located close to compact star forming knots, and those that are not. A good example of the former is SN\,2006gy, which exploded near the center of an early-type galaxy \citep{ofek2007,Smith2007}. SN\,2006gy exploded near a region showing signs of localized star formation. This is indicated by a nearby dust lane and compact \Ha\ emission interpreted as an H II region \citep{ofek2007,Smith2007}. However \cite{Jerkstrand2020} identify neutral iron lines in one of the spectra of SN\,2006gy, and argue it is consistent with a SN Ia embedded in a shell of circumstellar material, possibly contradicting a massive star origin. Similarly to the site of SN\,2006gy, the localized \Ha\ emission extending to the site of \ape\ is most likely an indication of a nearby H II region where the progenitor star could have been formed. SNe Ic are often associated with \Ha\ emission in their host galaxies, supporting a young progenitor population even when compared to other CCSNe types \citep{Anderson2015}. The same might be true for SNe Ic observed in elliptical galaxies. This would imply that the initial mass function of stars formed in H II regions in elliptical galaxies extends to massive ($\geq 8 M_{\odot}$) stars.

\subsection{Extended star-formation or faint and diffuse satellites?}
In addition to SNe occurring near sites of active formation, we identify a second group of SNe which occur in the outskirts of their elliptical hosts, with no signs of localized star formation. In all cases in which the background from the apparent host-galaxy was faint enough - we obtain strong limits on the presence of an underlying host galaxy. Limits on the luminosity of hypothetical faint hosts range from $-10\ \rm mag$ to $-12\ \rm mag$, putting all these hosts at the low end of the CCSN host-galaxy luminosity distribution as presented in \cite{Schulze2020}. 

\begin{figure*}
\centering
\includegraphics[width=2\columnwidth]{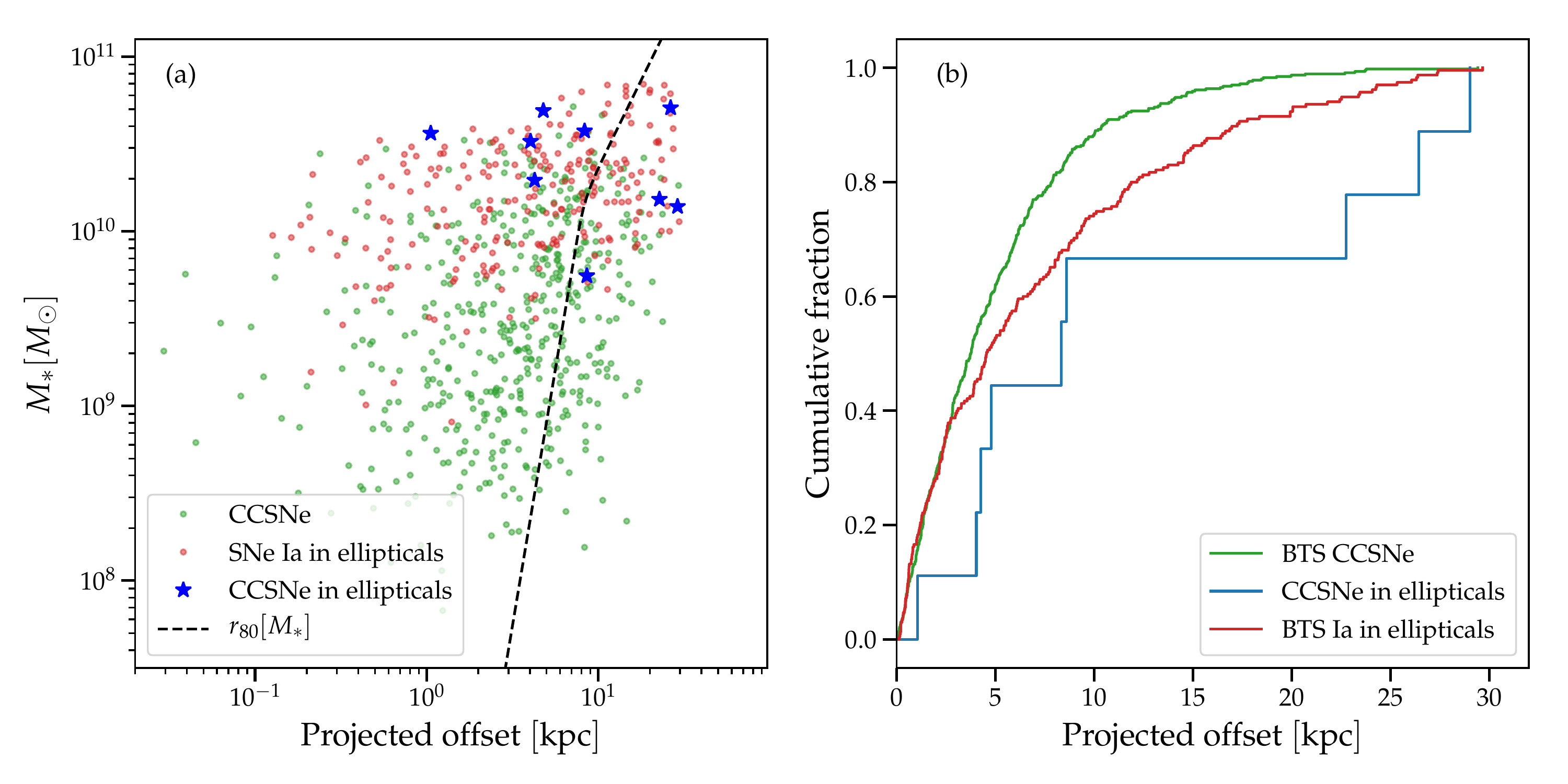}
\caption{Panel (a) shows the projected physical offsets of CCSNe in ellipticals (blue stars) and stellar mass of their host galaxies.  BTS Ia in ellipticals (red circles), and BTS CCSNe (green circles) are shown for comparison. The black dashed line represents the average 80\% light radius corresponding to the host-galaxy stellar mass. Panel (b) shows the cumulative distribution of the physical offsets for all three SN populations using the same color scheme.}  
\label{fig:offsets}
\end{figure*}

Next, we consider whether CCSNe are typically located at larger offsets from their elliptical host galaxies compared to the host-offset distribution of the general CCSN population, and compared to offsets of SNe Ia in elliptical galaxies. 
We begin by inspecting the comparison between the offset distributions of SNe Ia in ellipticals and our sample of CCSNe in ellipticals. The right panel of Fig. \ref{fig:offsets} shows the offset distribution of the BTS SNe Ia in ellipticals, CCSNe, and the combined sample of CCSNe in ellipticals. We show only SNe that are offset by less than 30 kpc (projected) or 90\arcsec, the limits of the automatic cross checks presented by \cite{perley2020}. We expect this to cover the vast majority of CCSNe, as the largest offsets in the (i)PTF sample were 37 kpc \citep{Schulze2020}. On the left panel of the same figure we show the physical offsets and the corresponding host-galaxy mass, along with the average 80\% light-radius, $r_{80}[M_{*}]$, derived using the low-redshift mass-size relation of \cite{Mowla2019}. \cite{Schulze2020} demonstrated (their Fig. 11) that the majority ($>85\%$) of CCSNe occur at distances smaller than $r_{80}$. 

The number of SNe in our sample is too low for significant results from a two-sample KS test, but we can formulate an alternative statistical test. We observe that $5\%$ of SNe Ia with elliptical host galaxies are detected at projected offsets larger than $22.5\ \rm kpc$. We define the null hypothesis as the case in which CCSNe in ellipticals come from the same offset distribution as SNe Ia: For a total of 9 SNe, we expect on average $\sim0.5$ SNe at offsets larger than $22.5\ \rm kpc$ (upper 5 \% of the Ia sample distribution). Assuming a Poisson distribution, the probability of finding three or more SNe in this range is then $1 \%$, so that the null hypothesis is rejected with a confidence of $99\%$. We note that this test is sensitive to the position of the threshold. For example, repeating this test for all SNe at distances larger than 17 kpc (upper 10 \% of the Ia sample distribution) will result in less significant results ($95\%$ rejection). Repeating this analysis with $r/r_{80}[M_{*}]$ gives similar results - the probability that 3 of 9 SNe have offsets larger than $>2\ r_{80}$ is less than $3\%$, assuming the offset distribution of SNe Ia with elliptical hosts. To conclude, we find tentative evidence that CCSNe are typically located at larger offsets from their host galaxies compared to SNe Ia in the same host-galaxy population. However, a larger sample is needed in order to reach a high statistical significance. \\

A possible explanation for the high offset of CCSNe in elliptical galaxies is a reduced detection efficiency at low offsets on top of the high surface brightness center of their elliptical hosts. \cite{Foley2015} suggested this option to explain the lack of Ca-rich SNe Ib at low offsets from their host-galaxies. \cite{Frohmaier2017} calculated the recovery efficiency of the PTF pipeline and found it was lower in regions of high surface brightness, but  \cite{Frohmaier2017} later found that this is not enough to explain the large offsets observed for Ca-rich transients \citep{Kasliwal2012}. ZTF data should be less susceptible to this bias than PTF was, since the new ZTF camera provides substantially higher image quality, and the optimal image subtraction of \citet{Zackay2016} provides much cleaner subtraction; we thus estimate this effect should be even weaker in our BTS sample. Still, a larger homogeneous sample is needed to quantify this effect for CCSNe in elliptical host galaxies. 

\indent An extended offset distribution for CCSNe in elliptical host galaxies compared to the progenitors of SNe Ia has significant implications for the formation of their progenitor stars. 

An offset between the locations of SNe Ia and CCSNe in ellitpicals could reflect an inside-out growth \citep[][]{SanchezBlazquez2007,perez2013} in some massive ellipticals. While the SNe Ia originate from an older stellar population closer to the center of the massive hosts, CCSNe could explode in regions where the host-galaxy is accreting gas from the IGM. This would be in agreement with the findings of \cite{Salim2012}, who found galaxy-scale star formation in UV-excess early-type galaxies can extend to large offsets. \cite{Gomes2016} found the extended star-formation in the periphery of early-type galaxies can take the form of faint ($24 \lesssim \mu_{r} (\rm mag/ \square \arcsec) \lesssim 26$) spiral features which  possibly indicate inside-out growth. However, such low-surface-brightness features are excluded for both SN\,2016hil and PS1-12sk, but are still an option for \fsh\, \uik\ and PTF10gqf. 

\indent It is possible that the SNe we observe do not originate from the ellipticals themselves, but from a different stellar population nearby. This option is supported by some evidence that the space between elliptical galaxies might not be completely empty. in addition to having a large offset distribution extending to regions with apparently no stellar populations \citep{Kasliwal2012}, Ca-rich SNe prefer group and cluster environments \citep{Lunnan2017}. Interestingly enough, PS1-12sk exploded in a bright cluster environment as noted by \cite{sanders2013}, who raised the possibility this might be evidence for star formation in galaxy cluster cooling flow filaments. Similarly, \fsh\ occurred in the compact group V1CG 538 \citep{Lee2017}.  

However, out of five object which do not coincide with their elliptical host galaxy, only two are possibly associated with a group or cluster environment. We consider this preliminary evidence that star-forming cooling flows are not the main channel of star formation near elliptical galaxies. \cite{Galyam2003} measured the fraction of cluster SNe Ia originating from an intergalactic stellar population (i.e. those which originate outside of galaxies) to be 20\%. In our sample, all SNe detected in group or cluster environment were offset from their putative host, indicating that this fraction is significantly different in CCSNe near ellipticals. However, a larger sample is needed to measure this fraction with high significance.  While not part of our sample, we also note that the SN II Abell399 11 19 0 \citep{Graham2012} occurred in a nearby galaxy cluster, and is not offset from its host galaxy.\\
 
\cite{Ruiz2014} explore a sample of 1000 ellipticals and find that the majority have small satellites (down to a mass ratio of 1:400) with an average projected distance of $\sim 59$ kpc from their parent galaxy, that contain 8\% of their stellar mass. However, such galaxies are excluded by our deep limits at the SN sites. Another possibility for the origins of offset CCSNe can be found in the work of \cite{Sedgwick2019}, who recently demonstrated that a population of star-forming Low-Surface-Brightness Galaxies (LSBGs) host many of the seemingly hostless SNe. These LSBGs can have significant star formation activity, like the nearby UGC 2162. This ultra-diffuse galaxy has a low (but not negligible) SFR of 0.01 \msunyr, but with a very low surface brightness of 24.4 $\rm mag/ \square \arcsec$ \citep{Trujillo2017}. This would translate to an integrated absolute magnitude of $-7\ \rm mag$ in the optical. Such a faint and extended object  would be difficult to detect when located on top of the diffuse emission at the outskirts of a massive elliptical at a redshift of 0.02. While such an ultra-diffuse host is ruled out in the case of PS1-12sk \citep{Hosseinzadeh2019}, it has not been excluded for SNe II in the outskirts of elliptical galaxies.

If the host galaxies of offset CCSNe are satellites of massive galaxies, they would constitute a small fraction of CCSNe in their typical host galaxies, and could erroneously be associated with nearby more luminous hosts. However, in elliptical galaxies which host much fewer CCSNe, such satellite galaxies could host a larger fraction of the observed SNe. To test this hypothesis, we select CCSNe from the BTS and PTF samples that occur at offset of $\geqslant15\ \rm kpc$ from massive ($\rm log(M[M_{odot}]) > 10$) galaxies. We manually examine the deepest publicly available survey data (if possible using the Legacy Survey or otherwise PS1 images) and exclude all cases where the SN occurred on or very close to pronounced structures that are part of the host galaxy, such as spiral arms extending to the SN location, or elevated ($>3 \sigma$) emission compared to the local background. We find the fraction of highly offset CCSNe occurring in elliptical compared to non-elliptical hosts is 0.33 (0.17-1.57; 95\% confidence interval). This value is consistent with the fraction of massive ellipticals of all massive ($\rm log(M[M_{\odot}]) > 10$) galaxies in the Galaxy Zoo sample - 0.33 (0.30-0.40; 95\% confidence interval). To reduce the selection bias, we selected a low-z ($0.025<z<0.032$) sample such that the redshift distributions of elliptical galaxies, spiral galaxies, and all galaxies (including those with irregular or uncertain morphological classifications) match.

We conclude that the most likely explanation for the highly-offset distribution of CCSNe near ellipticals is one of the following
\begin{itemize}

    \item a population of ultra-diffuse satellite star-forming galaxies near massive elliptical galaxies. The rate of highly offset CCSNe around massive ellipticals is hence similar to their rate around all massive galaxies. Due to the large number of CCSNe occurring in massive star-forming galaxies, the few offset events are unremarkable; but with very few events occurring in massive ellipticals, the offset population, perhaps arising from very faint satellites, stands out. Deeper observations of the location of our sample events are needed to establish this. 

    \item Some cases can be explained by extended and faint spiral features, or extended galaxy-scale star-formation due to accretion of gas from the IGM. The later would indicate inside-out growth of their elliptical host-galaxies. Future studies of the host-galaxies of the CCSNe in our sample can also confirm or exclude this possibility. 
    
\end{itemize}

\subsection{The star formation fraction in elliptical galaxies}
\begin{figure*}
\centering
\includegraphics[width=2\columnwidth]{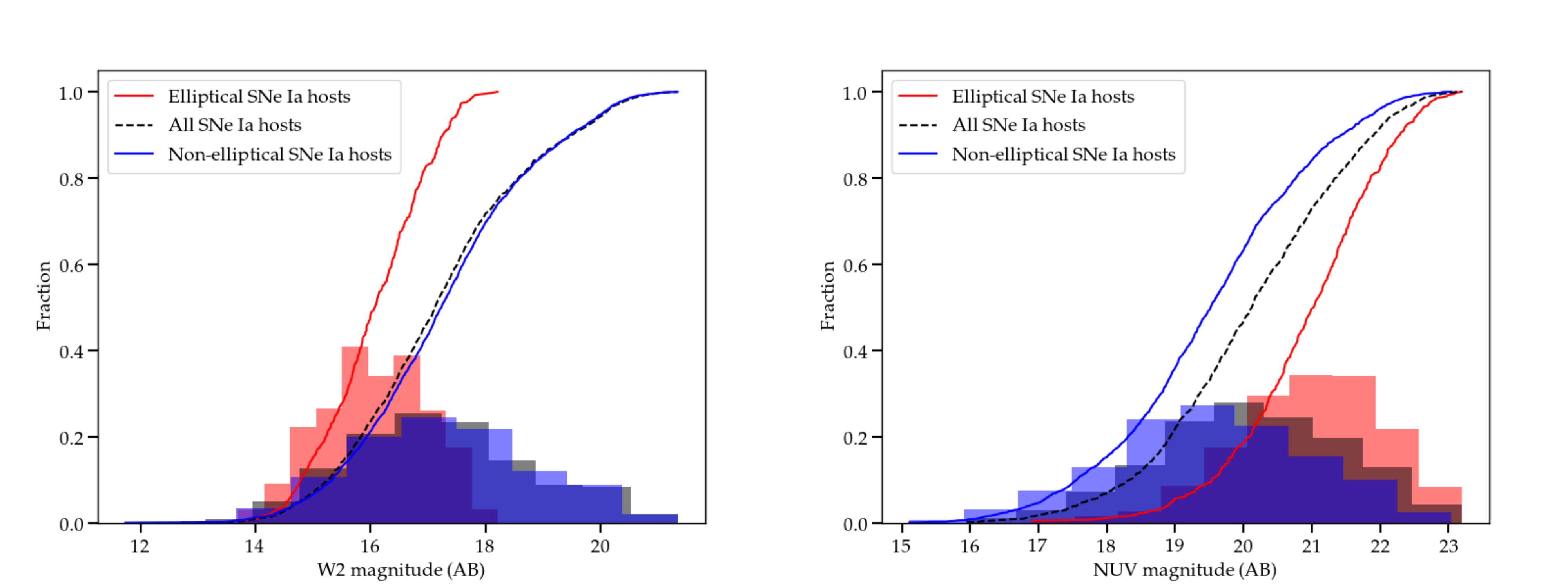}
\caption{W2 (left) and NUV (right panel) magnitude distribution for all BTS SNe Ia, BTS SNe Ia in elliptical galaxies and BTS SNe Ia in non-elliptical galaxies.}  
\label{fig:mag_hist}
\end{figure*}

We attempt to measure the formation rate of massive stars in elliptical galaxies through the fraction of CCSNe occurring in such environments in a fixed volume. The connection between the SFR and the CCSN rate has been previously established. \cite{Kennicutt1984} found a linear correlation between the SFR and the CCSN rate, and used it to estimate a lower mass limit for the progenitor mass of CCSNe. \cite{Botticella2012} assume a lower mass limit of 8 \msun\ and derive a SFR for a complete sample of galaxies in the local universe. Their estimate agrees with the SFR implied from their FUV luminosities. \cite{Maoz2011} used a sample of 119 CCSNe from the Lick Observatory SN search (LOSS; \citealt{Leaman2011}, \citealt{Li2011}) and estimated a rate of $0.01\pm0.002$ SNe per \msun. If the CCSN rate per \msun\ is fixed across galaxy type, the fraction of CCSNe occurring in elliptical galaxies should reflect the fraction of the total star formation in their host galaxy population. Recently, \cite{Schulze2020} provided additional support for this claim by showing that the SN host-galaxy mass distributions are consistent with those of star-forming galaxies weighted by their star-formation activity. \\
\indent When measuring the fraction of CCSNe in elliptical galaxies, the different colors of elliptical and spiral galaxies could create a selection effect especially when considering galaxies with both UV-optical and MIR color information. Most notably, the redder UV-optical color of elliptical galaxies makes them less likely to have archival \textit{GALEX} photometry compared to spirals at a similar optical brightness. However, our MIR photometry is not biased against ellipticals in the same way due to the brighter $W2$ magnitudes of massive ellipticals. $W3$ magnitudes distribute similarly for spirals and ellipticals. Figure \ref{fig:mag_hist} shows the magnitude distribution for the \textit{WISE}/W2  band and \textit{GALEX}/NUV band for all galaxies in the BTS Ia sample for elliptical and non-elliptical hosts. We illustrate that while ellipticals are less likely to be detected by \textit{GALEX} at larger distances due to their faint NUV brightness, they are more likely to be detected by \textit{WISE}. Thus, we derive estimates based on hosts with both UV-optical and MIR colors and based on hosts with MIR colors alone (using both definitions discussed in Sec. \ref{subsec:candidate}). A consistent result between both calculations will ensure that the selection effects in our samples are not strong. We also include a third estimate of the rate compared to all CCSNe, regardless of their host-galaxy association. The resulting fractions are:

\begin{itemize}
    \item $R_{CC}=\frac{2}{478}=0.4^{+0.5}_{-0.3}\%$ using both $W2-W3$ and NUV-\textit{r} color
    \item $R_{CC}=\frac{3}{888}=0.3^{+0.4}_{-0.1}\%$ based on our definition for ellipticals using only the $W2-W3$ color and compared to all CCSNe associated with host galaxies. 
    \item $R_{CC}=\frac{3}{959}=0.3^{+0.3}_{-0.1}\%$ based on our definition for ellipticals using only the $W2-W3$ color, and compared to all CCSNe regardless of their host-galaxy association. 
\end{itemize} 

Statistical errors reflect the exact binomial 68\% confidence interval on the reported values. We now compare the measured fraction of CCSNe in ellipticals ($\sim0.4\%$) with the fraction of star-formation ellipticals accounted for out of the total star formation produced by all galaxy types, calculated using the method of  \cite{Cortese2012} applied to UV and MIR observations. 

\indent We sum the total SFR of all Galaxy Zoo ellipticals which satisfy the color criteria in Sec. \ref{subsec:candidate}, and are within the range $0.025<z<0.032$, and compare it to the fraction of the total SFR in this redshift range in all galaxy types. We find that in this sample, the fraction of the total SFR produced by ellipticals is $1\pm0.01\%$ - slightly higher ($\sim 2 \sigma$) than the star formation inferred from the fraction of CCSNe in elliptical galaxies compared to all CCSNe. This could be explained due to additional UV emission by Active Galactic Nuclei (AGN) and old stellar populations \citep{Jura1982,Crocker2011}, or due to an incomplete sample  in the low-mass range of spiral galaxies (which host a significant fraction of CCSNe; \citealt{Taggart2021}). Alternatively, this difference could reflect a larger number of undetected SNe in elliptical compared to spiral galaxies due to different total extinction of the host galaxies, or due to a different rate of failed SNe in both environments.

\indent Our results are in significant tension with those of  \cite{Kaviraj2014}, who claim that early-type galaxies (elliptical and lenticular) account for $14\%$ of the star formation budget at $z<0.07$. The sample in the study by \cite{Kaviraj2014} is limited to bright $<16.8\ \rm mag$ galaxies, which account for 52\% of the total star formation in our general Galaxy Zoo sample, but only for 5\% of star formation in ellitpicals. Thus ellipticals are over-represented in their sample.  

\indent Our findings highlight the importance of CCSN rates as independent tracers of SFR across a large range of sSFR and galaxy brightness, while preventing the severe selection effects associated with the low end of the galaxy brightness distribution. The planned Rubin Observatory Legacy Survey of Space and Time (LSST; \citealt{Ivezic2019}) is expected to increase the rate of CCSN detections by an order of magnitude in upcoming years. This will increase the number of CCSNe detected in elliptical hosts by an order-of-magnitude,  allowing an in-depth study of the nature of star formation in and around elliptical hosts. However, we caution against using CCSNe as estimators of star formation without taking proper care to remove the significant contamination due to misclassified galaxies or SNe, or due to a false association of the SN with the host galaxy.

\indent \cite{dellavalle2005} previously estimated elliptical galaxies might host up to 3\% of CCSNe, based on a sample of SNe Ia in ellitpicals from the Asiago SN survey \citep{Cappellaro1999}. Our findings place more stringent constraints on the rate of CCSNe in ellipticals compared to those obtained in previous studies by an order of magnitude. Recently, \cite{Sedgwick2021} published a study of 36 CCSNe occurring in elliptical galaxies out of 421 photometrically-classified CCSNe isolated from the SDSS-II survey at $z<0.2$, suggesting that the fraction of CCSNe occurring in elliptical galaxies is significantly higher than the fraction we measure in this study ($\sim 8\%$ compared to our $\sim 0.4\%$). We point out that a population comprising of $10\%$ of CCSNe would stand out in the spectroscopically complete BTS, and conclude that this figure is most likely an overestimation due to a contamination of the sample by misclassified SNe. \cite{Sedgwick2021} define a confidently classified CCSN as having probability of $P_{Ia}<0.05$ for being a SN Ia. Since SNe Ia vastly outnumber CCSNe in elliptical hosts (in our sample, 240:2) the contamination of the CCSN sample due to falsely classified SNe Ia is significant, and likely dominates their sample.

\section{Summary and Conclusions}
\label{sec:conclusions}


In this study, we conducted a systematic search for elliptical host galaxies of CCSNe in the ZTF BTS - a spectroscopically complete survey. In Section \ref{sec:sample}, we outlined the selection process for the candidates. We identified elliptical galaxies based on their UV-optical and MIR colors, or based on the MIR color alone. A control sample of morphologically classified galaxies confirms that we achieve a good separation between elliptical and spiral galaxies. We further visually verified that the morphology of the host galaxies is elliptical, with no disc or bar structures.

\indent We identified three CCSNe that seem to be associated with elliptical galaxies. We presented their light curves,  spectra and host-galaxy photometry in Sec.\ref{sec:observations}, and analyzed our observations in Sec. \ref{sec:results}. \fsh\ and \uik\ are spectroscopically regular SNe II which exploded in the outskirts of massive ellipticals with no signs of an alternative underlying host. \uik\ shows typical spectra for SNe II during its photospheric phase, while \fsh\ shows an extended and boxy \Ha\ profile in its nebular spectrum, a signature of CSM interaction.  \ape\ is a SN Ic located in a massive elliptical near a compact star-forming region. Its \textit{g}-band light curve shows a late-time second peak  - a possible signature of extended CSM interaction, previously observed in other SNe Ic. Aside from this feature, a comparison to other SNe Ic shows it is a regular SN Ic and modelling a series of photospheric spectra reveals it had typical explosion properties. These, along with its location near a star-formation site, point to a massive star origin.

\indent  We proceed to combine these objects with 7 literature CCSNe satisfying our criteria, and analyzed the properties of the combined sample of SNe and their host galaxies. We derived stellar masses and SFRs for the hosts and compared these to a control sample of SNe Ia in elliptical galaxies and BTS CCSNe. Our analysis concludes that elliptical galaxies hosting CCSNe are broadly consistent with the general population of ellipticals. We discussed the implications of our results in Sec. \ref{sec:discussion}. We demonstrated that the peak \textit{r}-band absolute magnitudes of SNe II in ellipticals are consistent with those of a sample of SNe II from the BTS. We presented preliminary evidence that the offset distribution of CCSNe from their putative hosts extends to large offsets compared to both the BTS CCSN sample and a sample of SNe Ia in elliptical galaxies. However, this is not determined with high statistical significance, and verifying this requires a larger sample of events. We discussed possible reasons for the larger offset distribution and conclude that the most likely explanations are:
\begin{itemize}
    \item The CCSNe are hosted by a diffuse stellar population orbiting the nearby massive ellipticals. Since very few CCSNe occur in elliptical galaxies, a small population of CCSNe hosted by diffuse star-forming satellites would stand out. This is supported by the 1:3 ratio of highly offset CCSNe near spirals compared to elliptical galaxies - in agreement with ratio of massive spirals and ellipticals in the Galaxy Zoo sample. 
    \item The CCSNe originate from the nearby massive elliptical galaxy; either from faint peripheral spiral features, or due to extended galaxy-scale star formation due to gas accretion from the IGM. 
\end{itemize}

\indent Finally, we derived the fraction of CCSNe occurring in elliptical galaxies out of the general CCSN population and find it to be $0.4^{+0.5}_{-0.3}\%$ for SNe with a host galaxy associated with them, and $0.3^{+0.3}_{-0.1}\%$ when including host-less SNe. This is in slight tension ($\sim 2 \sigma$) with the fraction of star formation in morphologically classified ellipticals satisfying the same color criteria and calculated using traditional tracers (found to be $1\pm0.01\%$). However, this might be explained by systematics in the calibration of SFR relations in elliptical galaxies due to UV emission from old stellar populations, or due to selection effects biasing the galaxy sample in the low end of the galaxy mass distribution. We conclude that CCSNe can be useful as direct tracers of star formation in low sSFR environments, but caution that proper care must be taken to avoid contamination of both the SN sample (by misclassified SNe Ia) and the host sample (by star-forming galaxies).

\section{Acknowledgements }
AGY’s research is supported by the EU via ERC grant No. 725161, the ISF GW excellence center, an IMOS space infrastructure grant and BSF/Transformative and GIF grants, as well as The Benoziyo Endowment Fund for the Advancement of Science, the Deloro Institute for Advanced Research in Space and Optics, The Veronika A. Rabl Physics Discretionary Fund, Minerva, Yeda-Sela and the Schwartz/Reisman Collaborative Science Program;  AGY is the incumbent of the The Arlyn Imberman Professorial Chair.
NLS is funded by the Deutsche Forschungsgemeinschaft (DFG, German Research Foundation) via the Walter Benjamin program – 461903330.
This research has made use of the NASA/IPAC Extragalactic Database (NED),
which is operated by the Jet Propulsion Laboratory, California Institute of Technology,
under contract with the National Aeronautics and Space Administration (NASA).
Foscgui is a graphic user interface aimed at extracting SN spectroscopy and photometry obtained with FOSC-like instruments. It was developed by E. Cappellaro. A package description can be found at http://sngroup.oapd.inaf.it/foscgui.html.
This work is based on observations obtained with the Samuel Oschin Telescope
48-inch and the 60-inch Telescope at the Palomar Observatory as part of the
Zwicky Transient Facility project. ZTF is supported by the National Science
Foundation under Grant No. AST-1440341 and a collaboration including Caltech,
IPAC, the Weizmann Institute for Science, the Oskar Klein Center at Stockholm
University, the University of Maryland, the University of Washington,
Deutsches Elektronen-Synchrotron and Humboldt University, Los Alamos National
Laboratories, the TANGO Consortium of Taiwan, the University of Wisconsin at
Milwaukee, and Lawrence Berkeley National Laboratories. Operations are
conducted by COO, IPAC, and UW.
Based on observations from the Las Cumbres Observatory network.  The LCO team is supported by NSF grants AST-1911225 and AST-1911151
The Liverpool Telescope is operated on the island of La Palma by Liverpool John Moores University in the Spanish Observatorio del Roque de los Muchachos of the Instituto de Astrofisica de Canarias with financial support from the UK Science and Technology Facilities Council.
Partly based on observations made with the Nordic Optical Telescope, operated
at the Observatorio del Roque de los Muchachos, La Palma, Spain, of the
Instituto de Astrof\'isica de Canarias.
The SED Machine is based upon work supported by the National Science Foundation under Grant No. 1106171. The ZTF forced-photometry service was funded under the Heising-Simons Foundation grant \#12540303 (PI: Graham).
The Legacy Surveys consist of three individual and complementary projects: the Dark Energy Camera Legacy Survey (DECaLS; Proposal ID \#2014B-0404; PIs: David Schlegel and Arjun Dey), the Beijing-Arizona Sky Survey (BASS; NOAO Prop. ID \#2015A-0801; PIs: Zhou Xu and Xiaohui Fan), and the Mayall z-band Legacy Survey (MzLS; Prop. ID \#2016A-0453; PI: Arjun Dey). DECaLS, BASS and MzLS together include data obtained, respectively, at the Blanco telescope, Cerro Tololo Inter-American Observatory, NSF’s NOIRLab; the Bok telescope, Steward Observatory, University of Arizona; and the Mayall telescope, Kitt Peak National Observatory, NOIRLab. The Legacy Surveys project is honored to be permitted to conduct astronomical research on Iolkam Du’ag (Kitt Peak), a mountain with particular significance to the Tohono O’odham Nation.
BASS is a key project of the Telescope Access Program (TAP), which has been funded by the National Astronomical Observatories of China, the Chinese Academy of Sciences (the Strategic Priority Research Program “The Emergence of Cosmological Structures” Grant \# XDB09000000), and the Special Fund for Astronomy from the Ministry of Finance. The BASS is also supported by the External Cooperation Program of Chinese Academy of Sciences (Grant \# 114A11KYSB20160057), and Chinese National Natural Science Foundation (Grant \# 11433005)
This project has made use of data products from the Near-Earth Object Wide-field Infrared Survey Explorer (\textit{NEOWISE}), which is a project of the Jet Propulsion Laboratory/California Institute of Technology. NEOWISE is funded by the National Aeronautics and Space Administration.
Based on observations collected at the European Organisation for Astronomical Research in the Southern Hemisphere, Chile, as part of ePESSTO/ePESSTO+ (the extended Public ESO Spectroscopic Survey for Transient Objects Survey) under ESO programmes 199.D-0143,1103.D-0328,106.216C.001/007,106.216C.002/008 \& 106.216C.003/009.
This research has made use of the VizieR catalogue access tool, CDS,
 Strasbourg, France (DOI : 10.26093/cds/vizier). The original description of the VizieR service was published in 2000, A\&AS 143, 23
This work has made use of data from the Asteroid Terrestrial-impact Last Alert
System (ATLAS) project. ATLAS is primarily funded to search for near earth
asteroids through NASA grants NN12AR55G, 80NSSC18K0284, and
80NSSC18K1575; by products of the NEO search include images and catalogs from
the survey area. The ATLAS science products have been made possible through the
contributions of the University of Hawaii Institute for Astronomy, the Queen's
University Belfast, and the Space Telescope Science Institute.
TMB was funded by the CONICYT PFCHA / DOCTORADOBECAS CHILE/2017-72180113.
MN is supported by a Royal Astronomical Society Research Fellowship and by the European Research Council (ERC) under the European Union’s Horizon 2020 research and innovation programme (grant agreement No.~948381
MMK acknowledges generous support from the David and Lucille Packard Foundation. This work was supported by the GROWTH Marshal \citep{Kasliwal2019} developed as part of the GROWTH (Global Relay of Observatories Watching Transients Happen) project funded by the National Science Foundation under Grant No 1545949. MG is supported by the EU Horizon 2020 research and innovation programme under grant agreement No 101004719. L.G. acknowledges financial support from the Spanish Ministry of Science, Innovation and Universities (MICIU) under the 2019 Ram\'on y Cajal program RYC2019-027683 and from the Spanish MICIU project PID2020-115253GA-I00. T.-W.C. acknowledges the EU Funding under Marie Sk\l{}odowska-Curie grant H2020-MSCA-IF-2018-842471. ECK acknowledges support from the G.R.E.A.T research environment funded by {\em Vetenskapsr\aa det}, the Swedish Research Council, under project number 2016-06012, and support from The Wenner-Gren Foundations. L.G. acknowledges financial support from the Spanish Ministry of Science, Innovation and Universities (MICIU) under the 2019 Ram\'on y Cajal program RYC2019-027683 and from the Spanish MICIU project PID2020-115253GA-I00.

\bibliographystyle{apj} 
\bibliography{bibliograph.bib}

\end{document}